\documentclass[fleqn]{2017SCGE} 
\usepackage{color}
\usepackage{subfigure}
\usepackage{soul}
\soulregister\cite7 
\soulregister\cref7 
\newenvironment{figurehere} %
    {\def\@captype{figure}} 
    {}

\newcommand{\revi}{\color{black}}
\newcommand{\revii}{\color{black}}
\newcommand{\reviii}[1]{{\color{black}{#1}}}
\newcommand{\mathcolorbox}[2]{\colorbox{white}{$ #2$}}

\begin{document}

\ensubject{subject}

\ArticleType{Article}
\SpecialTopic{SPECIAL TOPIC: }
\Year{}
\Month{}
\Vol{}
\No{}
\DOI{}
\ArtNo{}
\ReceiveDate{}
\AcceptDate{}

\title{How will our knowledge of short gamma-ray bursts affect the distance measurement of binary neutron stars?}{How will our knowledge of short gamma-ray bursts affect the distance measurement of binary neutron stars?}

\author[1]{Minghui DU}{}
\author[1]{Lixin XU}{{lxxu@dlut.edu.cn}}

\AuthorMark{Du M H}

\AuthorCitation{Du M H, Xu L X}

\address[1]{Institute of Theoretical Physics, School of Physics, Dalian University of Technology, Dalian, 116024, P. R. China}


\abstract{Gravitational waves from binary neutron stars associated with short gamma-ray bursts have drawn considerable attention due to their prospect in cosmology. \reviii{For such events, the sky locations of sources can be pinpointed with techniques such as identifying the host galaxies.} However, the cosmological applications of these events still suffer from the problem of degeneracy between luminosity distance and inclination angle. To address this issue, a technique was proposed in previous study, i.e., using the collimation property of short gamma-ray bursts. Based on the observations, we assume that the cosine of inclination follows a Gaussian distribution, which may act as a prior in the Bayes analysis to break the degeneracy. This paper investigates the effects of different Gaussian priors and detector configurations on distance measurement and cosmological research. We first derive a simplified Fisher information matrix for demonstration, and then conduct quantitative analyses via simulation. By varying the number of third-generation detectors and the scale of prior, we generate four catalogs of 1,000 events. It is shown that, in the same detecting period, a network of detectors can recognize more and farther events than a single detector. Besides, adopting tighter prior and employing multiple detectors both decrease the error of luminosity distance. Also considered is the performance of a widely adopted formula in the error budget, which turns out to be a conservative choice in each case. As for cosmological applications, for the $\Lambda$CDM model, \reviii{500, 200, 600, and 300} events are required for the four configurations to achieve $1\%$ $H_0$ accuracy. With all 1,000 events in each catalog, $H_0$ and $\Omega_m$ can be constrained to (\reviii{$0.66\%$, $0.37\%$, $0.76\%$, $0.49\%$}), and (\reviii{0.010, 0.006, 0.013, 0.010}), respectively. The results of the Gaussian process also show that the gravitational wave standard siren can serve as a probe of cosmology at high redshifts.}
 


\maketitle


\begin{multicols}{2}
  \section{Introduction\label{Introduction}}
  Ever since the first observation of gravitational wave (GW) emitted by binary neutron star (BNS), known as GW170817~\cite{PhysRevLett.119.161101}, and the coincident short gamma-ray burst (SGRB) event GRB170817A~\cite{Goldstein:2017mmi}, the method of using gravitational wave standard siren (GWSS)~\cite{Schutz} together with electromagnetic (EM) counterpart~\cite{GBM:2017lvd} to study cosmology has come into reality. The GWs from compact binaries provide 
  \Authorfootnote
  direct measurements of the sources’ luminosity distances. The redshift of a GW source can be independently obtained via extra information from EM counterparts~\cite{Holz:2005df} or with statistical methods, such as the dark siren~\cite{MacLeod:2007jd, Chen:2017rfc,Fishbach:2018gjp}. Among various EM counterparts, SGRB, which is usually associated with BNS, appears especially useful. By identifying the host galaxy of SGRB, it is possible to determine the redshift and sky position of a GW source at the same time. Moreover, the collimation property of SGRB can also give a clue to the source's orientation. Although GW sources of other types are also reported possessing possible counterparts~\cite{Bhattacharya:2018lmw,Mukherjee:2020kki,Yi:2019rwo,Wang:2019bbk}, we are particularly interested in the cosmological applications of BNS-SGRB events.

  The second-generation GW detectors, e.g., advanced Laser Interferometer Gravitational-Wave Observatory (LIGO)-Hanford, advanced LIGO-Livingston, advanced Virgo, Kagra, and LIGO India, will soon operate with target sensitivity. Meanwhile, ground-based third-generation (3G) GW detectors, such as the Einstein Telescope (ET) in Europe~\cite{Punturo_2010,Sathyaprakash:2012jk} and Cosmic Explorer (CE)~\cite{2015PhRvD..91h2001D,Reitze:2019iox} in the US, as well as space interferometers like Tianqin in China, are also in the design phase. One of the key advantages of the 3G GW detectors is the extended sensitivity at low frequencies, which increases the detecting period of a single BNS event to several hours. Consequently, the time dependence of the detector tensor would become significant, and even an individual detector could be effectively treated as a network~\cite{Zhao:2017cbb, Yu:2020vyy}. With the improvement in sensitivity and completeness of the network, we can reasonably expect GWSS to become a powerful tool for cosmology, comparable to other probes, such as the cosmic microwave background or Type Ia supernova (SN) in the era of 3G GW detectors. \reviii{\cite{Maggiore:2019uih} offers an introduction to ET’s detection capabilities as well as potential astrophysical and cosmological applications in detail.}

  The cosmological applications of GW signals rely on determining source parameters, especially the luminosity distance.
  Intrinsic parameters, such as the component masses, can be measured with exquisite accuracy from the phase of GW~\cite{PhysRevD.47.2198}. \reviii{In the presence of counterparts such as SGRB, it is possible to precisely probe the sky position of the source by identifying the host galaxy. } Unfortunately, since luminosity distance degenerates with the inclination angle, using GWSS as an absolute distance probe requires a mechanism to break the degeneracy~\cite{Nissanke:2009kt}. Employing a network of detectors contributes to decreasing the uncertainty of distance~\cite{PhysRevD.49.2658}. In recent years, a customary practice is to neglect the distance--inclination correlation and account for the influence of correlation by multiplying a factor of 2~\cite{Tjonnie}. We argue that the validity of this method should be evaluated by quantitative analysis.

  Several recent studies have shown that information on SGRB signals may help break the degeneracy. The authors of~\cite{Fan:2017rkg} applied a Bayesian framework to 1,000 simulated joint detections by advanced LIGO and advanced Virgo to demonstrate that combined SGRB and GW observations could improve the estimations of progenitor distance and inclination angle. Besides, investigated in~\cite{Chen:2018omi} is the role of EM measurement in the analysis of GW data to improve the precision of luminosity distance, and hence that of the Hubble constant. This approach was tested on GW170817 and simulated events observed by the HLV and HLVJI networks. Another demonstration of the idea can be found in~\cite{Guidorzi:2017ogy}. Based on these pioneering studies, our research is dedicated to investigating how knowledge of SGRB affect the measurement of luminosity distance, and its further influence on cosmology, with both model-dependent and model-independent methods.

  In this paper, we adopt a method inspired by~\cite{Nissanke:2009kt,Fan:2017rkg,Chen:2018omi}. To solve the problem of degeneracy, we employ the propertie of SGRB in the form of a Gaussian prior on the inclination angle. The posterior distribution of parameters is then calculated using the Fisher information matrix (FIM). In the simplified regime where there is only one stationary interferometer, the FIM of luminosity distance and inclination can be analytically deduced, giving intuitive explanations to the problem of degeneracy and how it is broken by the prior. By varying the number of 3G GW detectors and the scale of prior, four catalogs of GW events are simulated to test the abovementioned pipeline of parameter estimation.

  We give a detailed description of our methods in \cref{Methodology}, including the detector response of 3G GW detectors, the manipulations of FIM based on SGRB prior, and the settings of simulations. Presented in \cref{Results} is the error analysis of simulated GW catalogs, followed by the application of mock data to the $\Lambda$CDM model and Gaussian process (GP). {\revi We further study the effects of altering the form of prior and star formation rate (SFR) model in~\cref{sec:FurtherComparisons}.} The concluding remarks are given in \cref{Conclusion}. In this paper, we adopt the natural units $G = c = 1$.

  \section{Methodology\label{Methodology}}
  \subsection{Gravitational wave detector response\label{detector response}}
  The detector response of GWs from coalescing compact binary is briefly summarized in this section. We calculate the coefficients of vectors and tensors with respect to an imaginary geocentric frame. Given a binary system at sky position $\bm{\hat{n}} = \bm{\hat{n}}\left(\theta, \phi\right)$ with the orientation of the angular momentum $\bm{\hat{L}}$, a pair of axes can be constructed as follows:
  \begin{equation}
      \bm{\hat{X}} = \frac{\bm{\hat{n}} \times \bm{\hat{L}}}{|\bm{\hat{n}} \times \bm{\hat{L}}|}, \quad
      \bm{\hat{Y}} = -\frac{\bm{\hat{n}} \times \bm{\hat{X}}}{|\bm{\hat{n}} \times \bm{\hat{X}}|},
  \end{equation}
  where $\bm{\hat{n}}$, $\bm{\hat{L}}$ are both unit vectors, and we define the inclination angle as $\iota = \arccos\left(\bm{\hat{n}} \cdot \bm{\hat{L}}\right)$. The pair $\{\bm{\hat{X}}, \bm{\hat{Y}}\}$ spans the projected orbital plane orthogonal to $\bm{\hat{n}}$, and $\bm{\hat{X}}$ and $\bm{\hat{Y}}$ are along the major and minor axes, respectively. By denoting the angle between $\bm{\hat{X}}$ and the orbit’s line of nodes as the polarization angle $\psi$, the coefficients of $\bm{\hat{X}}$ and $\bm{\hat{Y}}$ can be explicitly calculated ~\cite{PhysRevD.63.042003}.
  The GW tensor in transverse traceless (TT) gauge is the sum of two polarizations
  \begin{equation}{\label{GWtensor}}
    \bm{h}^{\rm TT}(t) = h_+(t)\bm{e}^+ + h_\times(t)\bm{e}^\times,
  \end{equation}
  where $\bm{e}^{+, \times}$ are the basis tensors of polarization, defined as
  \begin{equation}
      \bm{e}^+ = \bm{\hat{X}} \otimes \bm{\hat{X}} - \bm{\hat{Y}} \otimes \bm{\hat{Y}}, \quad
      \bm{e}^\times = \bm{\hat{X}} \otimes \bm{\hat{Y}} + \bm{\hat{Y}} \otimes \bm{\hat{X}}.
  \end{equation}
  Under incoming GW, the output of a ground-based Michelson interferometer (labeled $I$) is a timeseries:
  \begin{equation}{\label{strain}}
      h_I(t) = \sum_{ij} h^{\rm TT}_{ij}(t)D_{I, ij}(t),
  \end{equation}
  where the detector tensor of an interferometer with arms $\bm{\hat{l}}$ and $\bm{\hat{m}}$ is $\bm{D}_I = \left(\bm{\hat{l}} \otimes \bm{\hat{l}} - \bm{\hat{m}} \otimes \bm{\hat{m}}\right) / 2$, and the time-dependence of $\bm{D}_I$ comes from the rotation of the Earth. For simplicity, we assume that the Earth is a perfect sphere with radius $R$ and rotational angular velocity $\Omega$. For an interferometer at $\bm{\hat{r}} = \bm{\hat{r}}(\pi / 2 - \phi_I, \lambda_I = \lambda_{I0} + \Omega t)$ ($\lambda_{I0}$ and $\phi_I$ are the east longitude and north latitude, respectively), whose $\bm{\hat{l}}$ arm is oriented at angle $\gamma_I$ north of east and $\bm{\hat{m}}$ arm at $\gamma_I + \zeta_I$, the unit vectors $\bm{\hat{l}}$, $\bm{\hat{m}}$ can be expressed as
  \begin{align}
      &\bm{\hat{l}} = \cos\gamma_I \bm{e}^E_I + \sin \gamma_I \bm{e}^N_I, \nonumber \\
      &\bm{\hat{m}} = \cos(\gamma_I + \zeta_I) \bm{e}^E_I + \sin (\gamma_I + \zeta_I) \bm{e}^N_I,
  \end{align}
  where $\bm{e}^E_I$ and $\bm{e}^N_I$ are the unit vectors pointing to the east and north on the Earth’s surface.
  Combining \cref{GWtensor} and \cref{strain}, the time-domain detector response reads
  \begin{equation}
     h_I(t) = h_+(t)F_I^+(t) + h_\times(t)F_I^\times(t),
  \end{equation}
  where $F_I^{+, \times} = \Sigma_{ij} D_{I, ij} e^{+, \times}_{ij}$ are the antenna pattern functions. A coalescing BNS system is characterized by component mass $(m_1, m_2)$, luminosity distance $d_{\rm L}$, time and phase of coalescence $(t_c, \phi_c)$; thus, the mass ratio is $\eta = m_1m_2 / (m_1 + m_2)^2$, and the redshifted chirp mass is $\mathcal{M}_c = (1 + z)(m_1 + m_2) \eta^{3/5}$.

  The orbital frequency varies negligibly over a single GW cycle during the inspiral section of BNS coalescence, which makes it possible to compute the Fourier transform of $h_I$ with the stationary phase approximation~\cite{Zhao:2017cbb}.
  In this paper, we adopt the restricted post-Newtonian (PN) approximation and calculate the waveform to the 3.5 PN order ~\cite{Zhao:2017cbb}:
  \begin{align}{\label{Fourier waveform}}
    \tilde h_I(f) = & \ A Q_I f^{-7/6} \exp [i(2\pi ft_c - \pi/4 + 2\Psi(f/2) \nonumber \\
      &-\phi_{I, (2, 0)} + 2\pi f \bm{\hat{n}} \cdot \bm{\hat{r}} )],
  \end{align}
  where the $2\pi f \bm{\hat{n}} \cdot \bm{\hat{r}}$ term is included to account for the time of GW propagating from
  the Earth’s center to the detector. Other terms are defined as follows:
  \begin{align}
      &A = \sqrt{\frac{5\pi}{96}}\pi^{-7/6}\mathcal{M}_c^{5/6}d_{\rm L}^{-1}, \\
      &Q_I = \sqrt{(1 + {\cos}^2 \iota)^2 (F_I^+)^2 + 4{\cos}^2\iota (F_I^\times)^2}, \\
      &\Psi(f) = -\phi_c + \frac{3}{256\eta}\sum_{i = 0}^7\psi_i(2\pi Mf)^{i/3}, \\
      &\phi_{I, (2, 0)} = {\rm tan}^{-1}\left[-\frac{2{\cos}\iota F_I^\times}{(1+{\cos}^2\iota)F_I^+}\right],
  \end{align}
  {\revii where $d_{\rm L}$ and $\iota$ are the luminosity distance and inclination angle, respectively, and $M$ represents the redshifted total mass, defined as $M = (1 + z)(m_1 + m_2)$.}
  Detailed expressions for the PN coefficients $\psi_i$ can be found in~\cite{Sathyaprakash:2009xs}. Notably, the time-dependence of $F_I^{+, \times}$ should be converted to frequency-dependence through $F_I^{+, \times}(f) = F_I^{+, \times}(t_f)$ where $t_f = t_c - (5/256)\mathcal{M}_c^{-5/3}(\pi f)^{-8/3}$. This principle also applies to $\bm{\hat{r}}$.

  Three 3G detectors will be considered in the following research: ET in Europe, CE in Idaho, USA, and an assumed CE-like detector in New South Wales, Australia. The parameters $\{\lambda_{I0}, \phi_I, \gamma_I, and \zeta_I\}$ are listed in Table III of~\cite{2020arXiv201015202B} \reviii{and~\cref{tab:detectors} of this paper. Besides, we choose the sensitivity curves of ET and CE as ET-D and CE2-40-CBO, whose definitions can be found in~\cite{2020arXiv201015202B}.} We denote the network of ET and two CE-like detectors as ET + CE.
  \begin{table*}
    \begin{center}
      \renewcommand\arraystretch{1.5}
      \begin{tabular}{p{2cm}p{1.8cm}<{\centering}p{1.8cm}<{\centering}p{1.8cm}<{\centering}p{1.5cm}<{\centering}}
      \hline
      Detector & $\lambda_{I0}$ & $\phi_I$ & $\gamma_I$ & $\zeta_I$ \\
      \hline
      ET & 0.183338 & 0.761512 & 2.802430 & $\pi / 3$ \\
      CE (USA) & -1.969170 & 0.764918 &  0 & $\pi / 2$\\
      CE (AUS) & 2.530730 & -0.593412 & 0.785398 & $\pi / 2$\\
      \hline
      \end{tabular}
    \end{center}
    \caption{\reviii{Configurations of the 3G gravitational wave detectors, including the longitudes $\lambda_{I0}$, latitudes $\beta_I$, orientations $\gamma_I$, and angles between two arms $\zeta_I$. All angles are in radians. The locations of detectors do not represent candidates under active consideration.}\label{tab:detectors}}
  \end{table*}

  
  \subsection{Fisher information matrix \label{Fisher matrix}}
  The detector response of GW entirely depends on nine parameters: $\{ t_c$, $\phi_c$, $\eta$, $\mathcal{M}_c$, $\theta$, $\phi$, $\psi$, $\iota$, $d_{\rm L} \}$. Alternatively, one can use GW signal to constrain these parameters, among which we are particularly interested in $d_{\rm L}$.
  Suppose that we have a set of observational data $D$ and a model $H$ with $N$ parameters dubbed $\bm{\lambda}$. According to the Bayesian theorem, the posterior of $\bm{\lambda}$ is
  \begin{equation}{\label{Bayes}}
      p\left(\bm{\lambda} | D, H\right) = \frac{p\left(D | \bm{\lambda}, H \right) \
       p\left(\bm{\lambda} | H\right)}{p\left(D | H\right)},
  \end{equation}
  where $p\left(D | \bm{\lambda}, H \right)$ denotes the likelihood function, and $p\left(\lambda | H\right)$ is the prior. The normalization
  factor $p\left(D | H\right)$ can be calculated by requiring
  $\int d^N \bm{\lambda} p\left(\bm{\lambda} | D, H\right) = 1$. If we choose flat priors on all parameters, then the posterior is proportional to the likelihood function.

  In general, the result of full Bayesian analysis is more precise and rigorous than that of FIM, and the Bayesian method is already adopted in topics such as GWSS~\cite{Chen:2020zoq} and dark siren. While, it has been shown that the results of FIM agree with sophisticated Bayesian parameter estimation in the condition of high signal-to-noise ratio (SNR)~\cite{PhysRevD.88.084013, PhysRevD.91.104001}, which, as will be seen later, is always satisfied in our simulation.
  \reviii{The implementations of FIM in the context of GW parameter estimation can be found in recent research such as~\cite{Wang:2020dkc,Wang:2020xwn, Grimm:2020ivq}.}
  More importantly, we are especially concerned about the relationship between luminosity distance and inclination, and the simplified FIM of these two parameters can provide intuitive explanations to the problem of degeneracy and how it is broken by the prior. Therefore, in this paper, we will implement FIM as the primary approach of parameter estimation.

  Following~\cite{PhysRevD.49.2658}, the joint posterior distribution of parameters can be approximated by a multidimensional Gaussian form:
  \begin{equation}{\label{param pdf}}
      p\left(\Delta \bm{\lambda}\right) \propto \exp\left(-\frac{1}{2}\sum_{ij}\Delta \lambda_i \
      \Gamma_{ij} \Delta \lambda_j \right),
  \end{equation}
  where
  \begin{equation}{\label{Fisher}}
      \Gamma_{ij} = \sum_I^{N_d} \Gamma_{I, ij} = \sum_I^{N_d} \left(\frac{\partial \tilde h_I}{\partial \lambda_i} \vline \
      \frac{\partial \tilde h_I}{\partial \lambda_i}\right)
  \end{equation}
  is the FIM of a detector network comprising $N_d$ interferometers ($N_d = 3$ for ET, $N_d = 5$ for ET + CE) and $\Delta \bm{\lambda}$ represent the deviations of $\bm{\lambda}$ to their “true” values.
  The inner product of any two functions $\tilde a(f)$ and $\tilde b(f)$ is
  \begin{equation}
      \left(\tilde a \, \vline \, \tilde b\right) = 4\Re\left[\int_{f_{\rm lower}}^{f_{\rm upper}}df \frac{\tilde a^\ast (f) \tilde b(f)}{S_{h,I}(f)}\right],
  \end{equation}
  $S_{h,I}(f)$ being the one-sided noise power spectrum density (PSD) of the $I$th detector. We set $f_{\rm lower} = 1$Hz and $f_{\rm upper} = 2f_{\rm LSO} = 1/(6^{3/2}\pi M)$, sufficient to cover the inspiral section of BNS coalescence, where $M = (1 + z)(m_1 + m_2)$ is the redshifted total mass (see ~\cite{Abbott_2017,Freise:2009nz,Zhao:2010sz} for the PSDs of ET and CE).
  We will find it convenient to give the definition of SNR:
  \begin{equation}{\label{SNR}}
      \rho = \sqrt{\sum \rho_I^2} = \sqrt{\sum (\tilde h_I | \tilde h_I)}.
  \end{equation}
  According to \cref{Fourier waveform},
  \begin{equation}{\label{SNR expression}}
      \rho_I^2 = \frac{5\mathcal{M}_c^{5/3}}{24 \pi^{4/3}d_{\rm L}^2} \int_{f_{\rm lower}}^{f_{\rm upper}}\frac{Q_I^2 df}{f^{7/3}S_{h, I}(f)}.
  \end{equation}

  Once we have $\Gamma_{ij}$, it is straightforward to evaluate the accuracies of parameters, characterized by the covariant matrix $\Sigma = \Gamma^{-1}$. Thus, the root-mean-square (RMS) error of $\lambda_i$ is $\sigma_{\Delta \lambda_i} = \sqrt{\Sigma_{ii}}$, and the covariance between two parameters $\left(\lambda_i, \lambda_j\right)$ is $c_{ij} = \Sigma_{ij} / \sqrt{\Sigma_{ii}\Sigma_{jj}}$.

  In the context of BNS-SGRB event, the sky position $(\theta, \phi)$ of the GW source can be pinpointed by techniques such as identifying the host galaxy. Moreover, the mock data challenge of the 3G GW detector~\cite{PhysRevD.86.122001} indicates that the mass parameters can be constrained to great accuracy. Since our interest is on the determination of luminosity distance, parameters such as $t_c$ and $\phi_c$, which only appear in the expression of phase, only have negligible impact. These considerations leave only $(\psi, \iota, d_{\rm L})$ to be determined.
  Further investigations on the parameter space~\cite{Nissanke:2009kt,PhysRevD.49.2658,PhysRevD.74.063006} have shown that the correlation between $d_{\rm L}$ and $\iota$ is overwhelmingly intensive; thus, we focus only on the covariant matrix of $d_{\rm L}$ and $\iota$ and assume that other parameters are perfectly constrained.

  It then comes down to calculating the derivatives of $\tilde h_I(f)$. Below are some relevant expressions, and the derivatives with respect to other parameters can be found in~\cite{Feng:2019wgq, Berti:2004bd}:
  \begin{align}
      &\frac{\partial \ln \tilde h_I}{\partial d_{\rm L}} = -\frac{1}{d_{\rm L}}, \label{derivativeD} \\
      &\frac{\partial \ln \tilde h_I}{\partial v} = \frac{1}{Q_I}\frac{\partial Q_I}{\partial v}
      - i\frac{\partial \phi_{I, (2, 0)}}{\partial v} \nonumber \\
      &\quad \quad \ \ \ = \frac{2v\left(1 + v^2\right)\left(F_I^+\right)^2 + 4v\left(F_I^\times\right)^2}{Q^2_I}
      + i\frac{2\left(1 - v^2\right)F_I^+F_I^\times}{Q_I^2} \nonumber \\
      &\quad \quad \ \ \ \equiv v_{1, I} + i v_{2, I}, \label{derivativev}
  \end{align}
  where we have replaced the parameter $\iota$ by $v = \cos\iota$ for convenience, and the real and imaginary parts of $\partial \ln \tilde{h}_I / \partial v$ are abbreviated as $v_{1,I}$ and $v_{2, I}$, respectively. Substituting \cref{derivativeD} and \cref{derivativev} into \cref{Fisher}, we have
  \begin{align}{\label{Fisher_dv}}
      &\Gamma_{I, d_{\rm L}d_{\rm L}} = \frac{\rho_I^2}{d_{\rm L}^2}, \nonumber \\
      &\Gamma_{I, d_{\rm L}v} = \Gamma_{I, vd_{\rm L}} =-\frac{4}{d_{\rm L}}\int \
      \frac{v_{1,I} |\tilde h_I|^2 df}{S_{h, I}}, \nonumber \\
      &\Gamma_{I, vv} = 4\int\frac{\left(v_{1,I}^2 + v_{2,I}^2\right)|\tilde h_I|^2 df}{S_{h, I}}.
  \end{align}
  Generally, $v_{1, I}$ and $v_{2, I}$ are frequency-dependent due to the rotation of Earth, and fixing the Earth’s orientation would increase resulting errors. However, the form of FIM would become quite simple under this assumption:
  \begin{equation}
      \Gamma_I = \left[
          \begin{array}{cc}
              \rho^2 / d_{\rm L}^2 & -v_{1,I} \rho^2 / d_{\rm L}  \vspace{0.1cm}\\
              -v_{1,I} \rho^2 / d_{\rm L} & \left(v_{1,I}^2 + v_{2,I}^2\right) \rho^2
          \end{array}
      \right],
  \end{equation}
  where the elements are arranged in the order $\left(d_{\rm L}, v\right)$.
  Further, in the case where there is only one interferometer ($N_d = 1$), dropping the suffix $I$, the inverse of FIM is as follows:
  \begin{equation}
      \Sigma = \left[
          \begin{array}{cc}
              \left(v_1^2 + v_2^2\right)d_{\rm L}^2 / v_2^2 \rho^2 & v_1 d_{\rm L} / v_2^2 \rho^2 \vspace{0.1cm}\\
              v_1 d_{\rm L} / v_2^2 \rho^2 & 1 / v_2^2 \rho^2
          \end{array}
      \right].
  \end{equation}
  Next, we will illustrate our technique of handling FIM under these simplifications and perform qualitative analyses. Whereas the full expression~\cref{Fisher_dv} will be used in the simulation (\cref{simulation}).

  The SGRBs are believed to be strongly beamed phenomena~\cite{Abdo2009FermiOO,Nakar,Rezzolla}, which was confirmed by recent investigations on GRB 170817A. {\revii From theoretical expectation, it is usually assumed that for SGRB induced by BNS coalescence, the viewing angle of SGRB is identical to the inclination angle $\iota$.} ~\cite{10.1093/mnrasl/sly061} proposed that $\iota$ follows Gaussian jet profile with mean value 0 and standard deviation $\sigma_\iota = 0.057^{+0.025}_{-0.023}$ rad, while~\cite{Howell:2018nhu} constrained $\sigma_\iota = 4.7^{+1.1}_{-1.1}$ deg. In other previous research~\cite{Nissanke:2009kt,PhysRevD.74.063006}, it was assumed that SGRB is confined within $25^\circ$.  As a result, the behavior of $\Sigma$ near $v = \cos \iota \rightarrow 1$ (face-on) is the most relevant.

  Since face-on and face-off are equivalent, we focus only on positive values of $v$. It is easy to figure that when $v \rightarrow 1$, $v_1 \rightarrow 1$ and $v_2 \rightarrow 0$; thus, the errors of $d_{\rm L}$ and $v$ (dubbed $\sigma_{\Delta d_{\rm L}}$ and $\sigma_{\Delta v}$) diverge, and the correlation $c_{d_{\rm L}v} = v_1 / \sqrt{v_1^2 + v_2^2}$ approaches 1, meaning that $d_{\rm L}$ and $v$ are intensively degenerate. Moreover, in the limit of $v = 1$, $\Gamma$ is singular so that $\Sigma$ cannot be well-defined, despite the fact that SNR actually increases for larger $v$.

  To address this issue, several methods have been proposed.   In recent years, a customary practice in the distance measurement of GW, as depicted by~\cite{Tjonnie}, is neglecting the correlation between $d_{\rm L}$ and $v$, calculating the error of $d_{\rm L}$ at $v = 1$ as $d_{\rm L} / \rho$, and accounting for the influence of varying $v$ with a factor of 2 (i.e., $\sigma_{\Delta d_{\rm L}} = 2 d_{\rm L} / \rho$, dubbed $\sigma_{\Delta d_{\rm L},0}$ hereafter)~\cite{Belgacem:2019tbw,Du:2018tia,Cai:2016sby,Zhao:2010sz}. We argue that this is different from considering the correlation at the first place; thus, the validity of using $\sigma_{\Delta d_{\rm L},0}$ should be assessed.

   \subsection{Short gamma-ray burst prior \label{prior}}
  In this section, we attempt to solve the problem of degeneracy by considering a nonuniform prior on $v$, assuming that we have knowledge about the properties of SGRB before detection. {\revii As stated in~\cref{Fisher matrix}, SGRB should be preferentially beamed along the orbital momentum axis of BNS.} Following~\cite{Nissanke:2009kt, PhysRevD.74.063006}, the prior on $v$ takes a Gaussian form with standard deviation $\sigma_v$:
  \begin{equation}{\label{vprior}}
      p_v(v) \propto \exp\left[-\frac{\left(1 - v\right)^2}{2\sigma_v^2}\right].
  \end{equation}
{\revii This is a pessimistic scenario where the inclination angle is determined with an accuracy of, e.g., $\sim$5 deg at $1\sigma$~\cite{10.1093/mnrasl/sly061, Howell:2018nhu}. Moreover, $\sigma_v$ is a global parameter so that a single value can be used to describe a whole population of astrophysical events that are uncorrelated.} Combining \cref{Bayes}, \cref{param pdf}, and \cref{vprior}, the posterior distribution of parameters is as follows:
  \begin{equation}{\label{posterior}}
      p(\Delta \bm{\lambda}) \propto \exp\left[-\frac{1}{2}\sum_{ij}\Delta \lambda_i \
      \Gamma_{ij} \Delta \lambda_j - \frac{\left(1 - v\right)^2}{2\sigma_v^2}\right],
  \end{equation}
  where $\bm{\Delta \lambda} = \left(\Delta d_{\rm L}, \Delta v\right)$.

  In general, the distribution given by \cref{posterior} is non-Gaussian due to the “true” value of $v$ is usually not exactly 1. However, we do not expect the errors to deviate from the face-on case considerably. As a schematic demonstration, we first analyze the situation where the “true” value of $v$ is 1. In that case, $p(\Delta \bm{\lambda})$ returns to the two-dimensional Gaussian form.
  Therefore, we can still describe the parameter accuracies in the framework of FIM. For the simplified regime described in \cref{Fisher matrix}, we have
  \begin{equation}{\label{simplified fisher}}
  \Gamma = \left[
      \begin{array}{cc}
          \rho^2 / d_{\rm L}^2 & -\rho^2 / d_{\rm L} \vspace{0.1cm} \\
          -\rho^2 / d_{\rm L} & \rho^2 + 1/\sigma_v^2
      \end{array}
  \right],
  \Sigma = \left[
      \begin{array}{cc}
          d_{\rm L}^2 / \rho^2 + d_{\rm L}^2 \sigma_v^2  & d_{\rm L} \sigma_v^2  \vspace{0.1cm} \\
          d_{\rm L} \sigma_v^2 & \sigma_v^2
      \end{array}
  \right].
  \end{equation}
  Interestingly, $\sigma_{\Delta d_{\rm L}} / d_{\rm L} = \sqrt{1 / \rho^2 + \sigma_v^2}$ gives an estimation comparable to $\sigma_{\Delta d_{\rm L},0} = 2 d_{\rm L} / \rho$, and the catastrophic error near $v = 1$ is avoided. Besides, $d_{\rm L}$ is better constrained when $\rho$ is larger or $\sigma_v$ is smaller, consistent with intuition. Also notable is that the correlation $c_{d_{\rm L}v} = \sigma_v / \sqrt{1 / \rho^2 + \sigma_v^2}$ is always smaller than 1.

  {\revi In the simplified regime, we further consider the situation where $v \neq 1$ and investigate the effect of varying $v$. To this end, a bunch of events detected by an interferometer with ET-D sensitivity is simulated, with the mass parameters of GW170817 ($\mathcal{M}_c = 1.188 M_{\rm sun}, m_2 / m_1 = 0.85$), located at redshifts 0.5, 1, 2, and 5. For reasons given below, two values of $\sigma_v$ (0.05 and 0.003) are adopted. When $v \neq 1$, \cref{simplified fisher} cease to be valid, and the FIM depends on $F^{+/ \times}$, hence the angles $\theta$, $\phi$, $\psi$; thus, we average $\sigma_{\Delta d_{\rm L}}$ over these parameters.

  \begin{figurehere}
    \centering
    \hspace{-0.5cm}
    \includegraphics[width=0.48\textwidth]{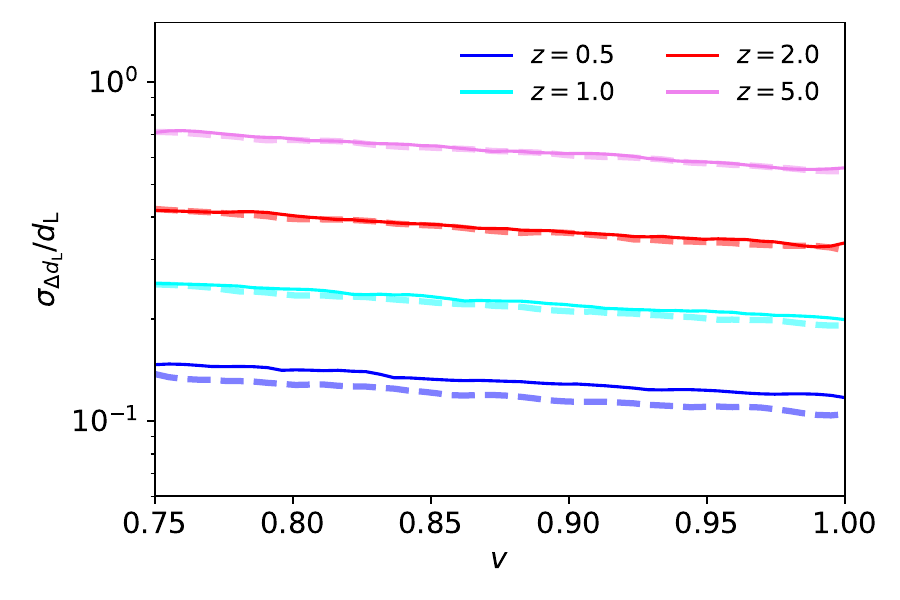}
    \vspace{-0.3cm}
    \caption{The relative errors of $d_{\rm L}$ as functions of $v$ at redshifts 0.5, 1, 2, and 5. The solid and dashed lines correspond to the results of $\sigma_v = 0.05$ and $\sigma_v = 0.003$, respectively.\label{fig:sigma_vs_z}}
    \vspace{0.2cm}
  \end{figurehere}

  Shown in~\cref{fig:sigma_vs_z} are the relative errors of $d_{\rm L}$ as functions of $v$, with $v$ ranging from $0.75$ to 1, equivalent to the $5\sigma$ range for $\sigma_v = 0.05$. The solid and dashed lines correspond to the results of $\sigma_v = 0.05$ and $\sigma_v = 0.003$, respectively. In each case, the error of $d_{\rm L}$ decreases as $v$ approaches 1. Moreover, with tighter prior on $v$, one can estimate $d_{\rm L}$ more accurately, and the influence of $\sigma_v$ becomes insignificant for distant events whose SNRs are relatively lower.}

  Notably, \cref{simplified fisher} merely serves to provide a qualitative illustration, and we will use the full expression \cref{posterior} in simulation. The error of luminosity distance is calculated by integrating over the entire distribution, following the standard definition in statistics.

  In this paper, two values of $\sigma_v$ will be considered: $\sigma_v = 0.05$, consistent with~\cite{Nissanke:2009kt,PhysRevD.74.063006} and $\sigma_v = 0.003$, equivalent to the 1$\sigma$ range of~\cite{Howell:2018nhu, Belgacem:2019tbw}. The first value seems rather large compared with the observation of GRB 170817A~\cite{Howell:2018nhu}, yet it can act as an upper bound for future observations. Although future SGRB observations may give $\sigma_v$ other than these two values, our research give a clue on the changes of results when $\sigma_v$ varies.

  The overall performance of our method depends on the distribution of source parameters. Details are presented in~\cref{simulation}.
  \begin{figure*}
    \centering
    \subfigure{\includegraphics[width=0.45\textwidth]{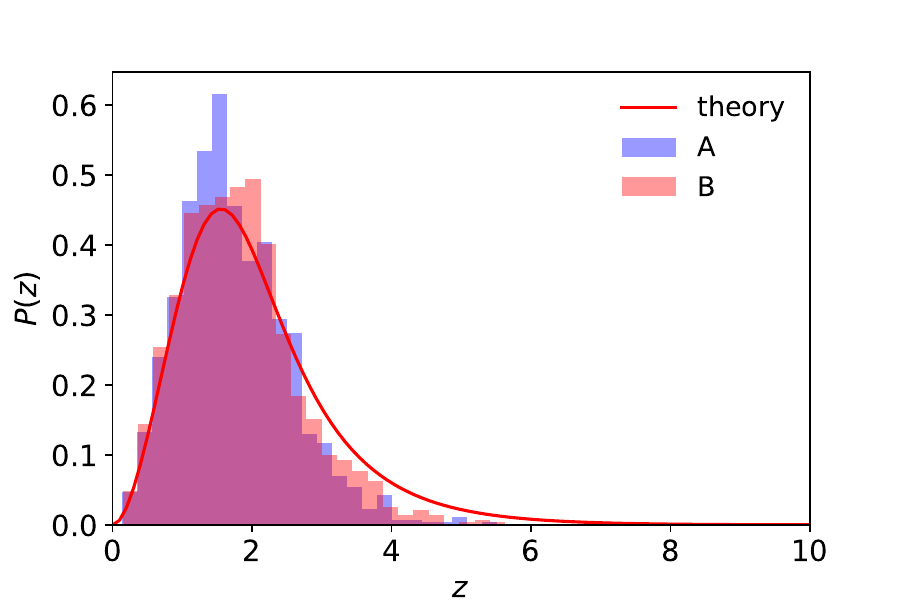}}
    \subfigure{\includegraphics[width=0.45\textwidth]{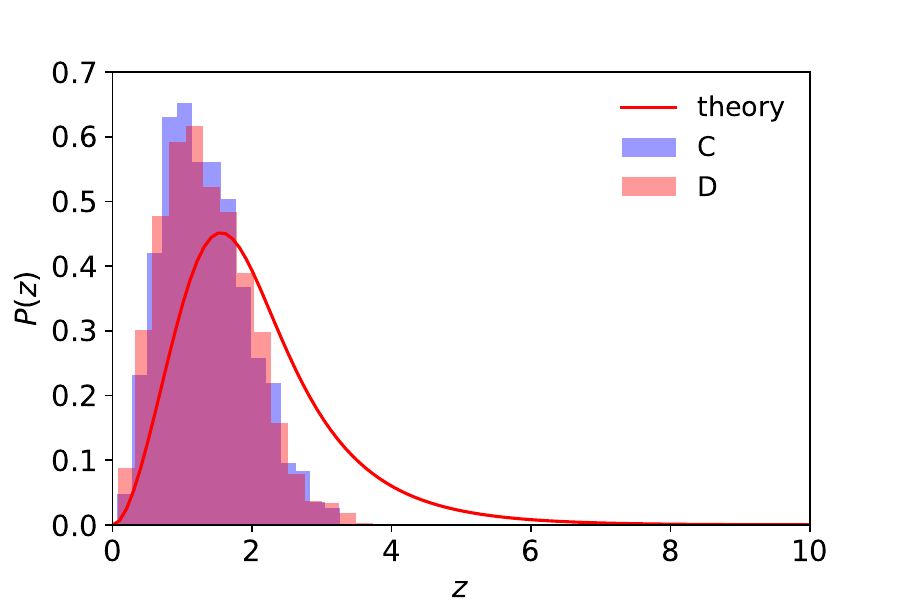}}
    \vspace{-0.2cm}
    \caption{The redshift distributions of simulated gravitational wave sources. $P(z)$ denotes the distribution of sources, normalized in the redshift range $z \in (0, 10)$. Shown in the left and right panels are the events detected by 3G GW detector network (ET + CE) and single detector (ET), respectively. In each panel, the histogram in blue (red) corresponds to $\sigma_v = 0.05 \ (\sigma_v = 0.003)$. The red curve with label “theory” is plotted according to \cref{redshift distribution}.\label{z_distribution}}
  \end{figure*}

  \subsection{Simulation\label{simulation}}
  The fiducial cosmological model is flat $\Lambda$CDM with parameters $H_0 = 67.8 {\rm km\ s^{-1}\ Mpc^{-1}}$ and $\Omega_m = 0.308$; thus, we have
  \begin{equation}{\label{dl_lcdm}}
      d_{\rm L}(z) = \frac{1 + z}{H_0}\int_0^z \frac{dz^\prime}{\sqrt{\Omega_m(1 + z^\prime)^3 + 1 - \Omega_m}}.
  \end{equation}
  The redshifts of sources are drawn from the distribution
  \begin{equation}{\label{redshift distribution}}
      R_z(z) = \frac{R_{\rm merge}(z)}{1+z}\frac{dV(z)}{dz}
      = \frac{4 \pi d_{\rm L}^2(z) R_{\rm merge}(z)}{(1 + z)^3 H(z)},
  \end{equation}
  where the second equality comes from the definition of comoving volume, and $H(z)$ is the Hubble parameter at redshift $z$. The merger rate of BNS $R_{\rm merge}(z)$ is calculated in the source frame, which is the convolution of neutron star (NS) formation rate $R_{\rm f}$ and a time-delay distribution $P_{\rm d}(t_{\rm d})$~\cite{Chen:2018rzo,Belgacem:2019tbw}:
  \begin{equation}
      R_{\rm merge}(z) = \int_{t_{\rm min}}^{t_{\rm max}}R_{\rm f}\left[t(z) - t_{\rm d}\right]\ P_{\rm d}\left(t_{\rm d}\right)dt_{\rm d},
  \end{equation}
  where $t_{\rm d}$ is the time between the formation and merger. By neglecting the lifetime and mass variation of the progenitor star~\cite{Schaerer:2001jc}, $R_{\rm f}$ follows the SFR
  \begin{equation}\label{eq:SFR}
      {\rm SFR}(z) = k\frac{a \exp[b(z-z_{\rm m})]}{a-b+b \exp[a(z-z_{\rm m})]}
  \end{equation}
  with parameters of the “Fiducial + PopIII” model~\cite{Dvorkin:2016wac}. Further, we assume
  $P_{\rm d}(t_{\rm d}) \propto t_{\rm d}^{-1}$ for $t_{\rm min} = 20$Myr and
  $t_{\rm max} = t_{\rm H}$, $t_{\rm H}$ is the Hubble time.

  As for other parameters, the sky location $\bm{\hat{n}} = \bm{\hat{n}}(\theta, \phi)$, polarization $\psi$, time and phase of coalescence $(t_c, \phi_c)$ are sampled from uniform distributions, and $v$ is drawn from \cref{vprior}. The component masses of BNS follow the Gaussian distribution $N(1.33, 0.09)M_{\rm sun}$, according to~\cite{LIGOScientific:2018mvr}. Besides, we classify a GW event as detectable if the total SNR exceeds the threshold of $\rho_{\rm threshold} = 12$ for ET + CE or 8 for ET. We set a lower threshold for ET to not limit its observation depth, making the population detected by ET and ET + CE comparable. These thresholds are widely adopted in the literature.

  Notably, current SGRB observations accumulate at only low redshifts, which is partly because the apparent luminosity of SGRB decreases with distance, and an event has to exceed the flux limit of the detector to be detected. However, from the theoretical perspective, coalescing BNS can be accompanied by SGRB provided proper conditions are satisfied. Therefore, we optimistically assume that SGRB counterparts at redshifts higher than the range of current observation can be detected in the future. We believe that with a reliable prediction of the SGRB detection rate, the analysis of our study would become more practical, while this is beyond the scope of this paper.

  We do not intend to predict the exact detection rate of BNS-SGRB events, given that we only have very few observations to date. Contrariwise, we will generate a catalog comprising 1,000 events (for each configuration), and for every single event, the detecting criteria of both GW and SGRB are satisfied. This requires that more than 1,000 sources should be simulated, and then the rate of acceptance is defined as 1,000 divided by the number of all (qualified or unqualified) sources. {\revi For current SGRB observation by Swift, the number of SGRBs with redshift measurement is less than 10 per year. However, the authors of~\cite{Belgacem:2019tbw} simulated GW detection with an ET + CE + CE network, assuming that a THESEUS-type ~\cite{THESEUS:2017qvx, THESEUS:2017wvz, 2018MmSAI..89..205S} satellite will be used for coincidence searches, and they obtained 907 GW-SGRB joint detections in 10 years of data collection if the mass distribution of BNS is Gaussian (as is assumed in our study). Therefore, we consider it reasonable to expect 1,000 joint events with the help of future THESEUS telescope.} Next, we will study how many events are sufficient for the target accuracy required by cosmological research. Four configurations will be considered in this paper: \{A: ET + CE /$\ \sigma_v = 0.05$, B: ET + CE /$\ \sigma_v = 0.003$, C: ET /$\ \sigma_v = 0.05$, D: ET /$\ \sigma_v = 0.003$\}.

  \begin{figure*}
    \centering
    \hspace{-0.3cm}
    \subfigure{\includegraphics[width=0.45\textwidth]{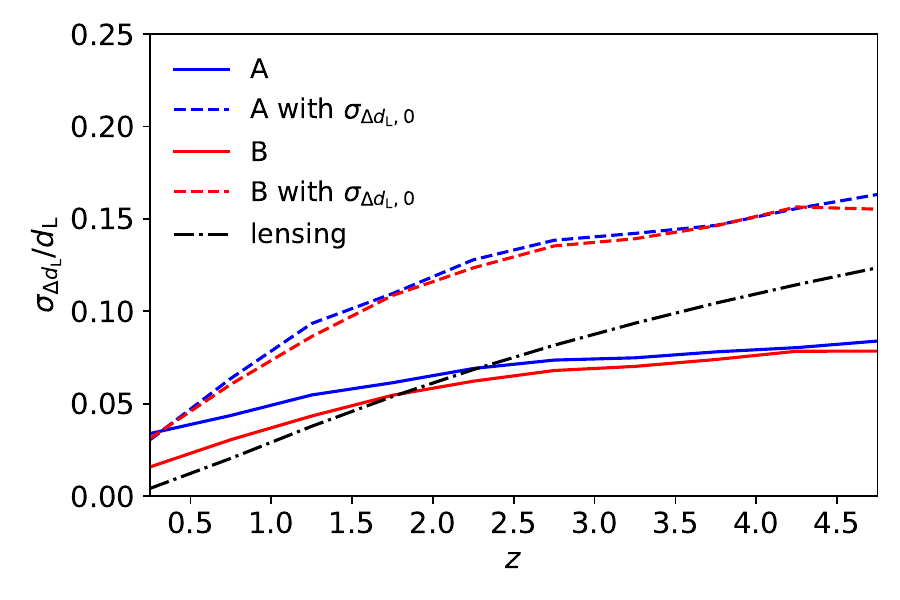}}
    \subfigure{\includegraphics[width=0.45\textwidth]{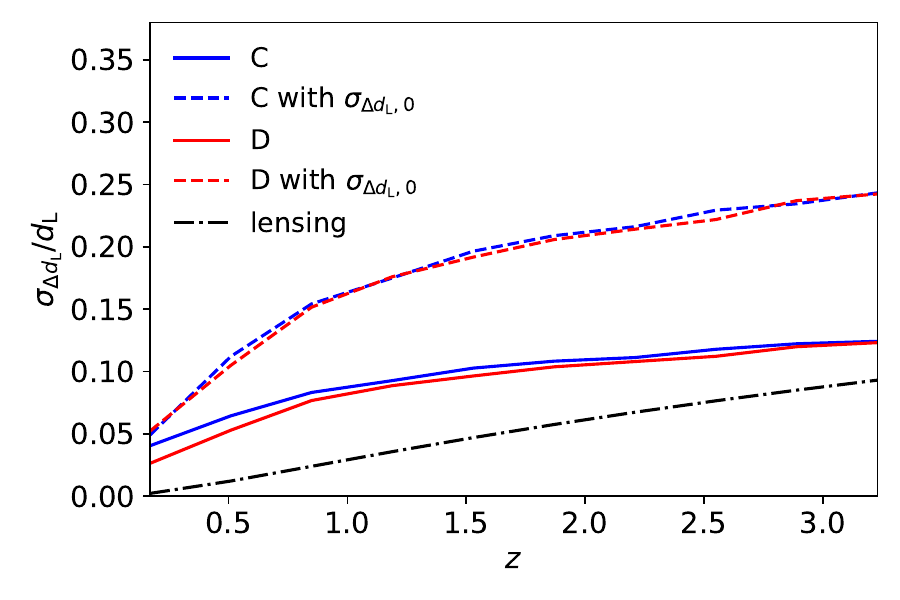}}
    \caption{\reviii{Average fractional errors of $d_{\rm L}$ in redshift bins. The results of catalogs A and B are plotted in the left panel, and those of C and D are shown in the right one. In both panels, the blue (red) curves correspond to $\sigma_v = 0.05 \  and (\sigma_v = 0.003)$, respectively. For comparison, $\sigma_{\Delta d_{\rm L},0} / d_{\rm L}$ for each configuration is also shown, and the black dash-dotted lines represent the errors induced by lensing.} \label{error_z}}
  \end{figure*}

  \begin{figure*}
    \centering
    \vspace{-1.5cm}
    \subfigure{\includegraphics[width=0.46\textwidth]{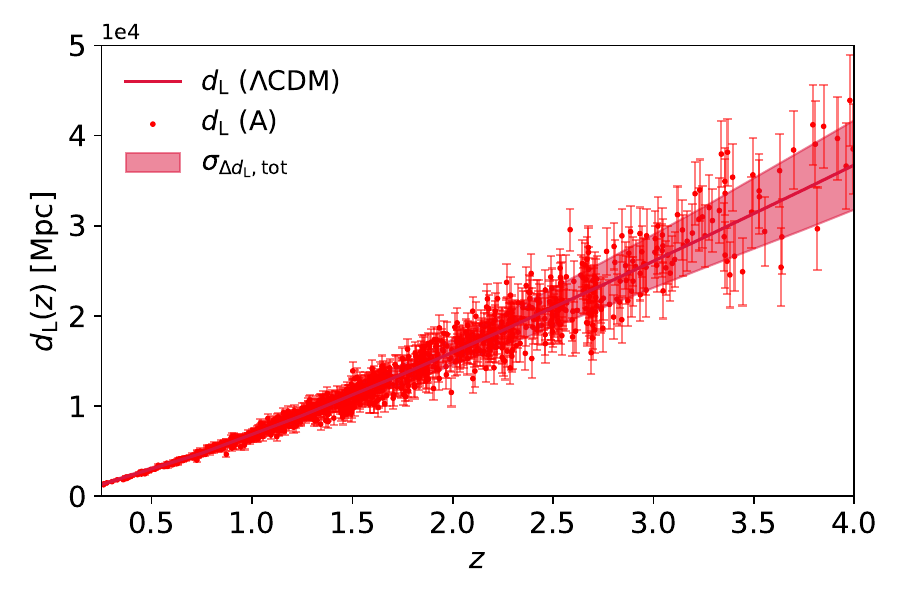}}
    \subfigure{\includegraphics[width=0.46\textwidth]{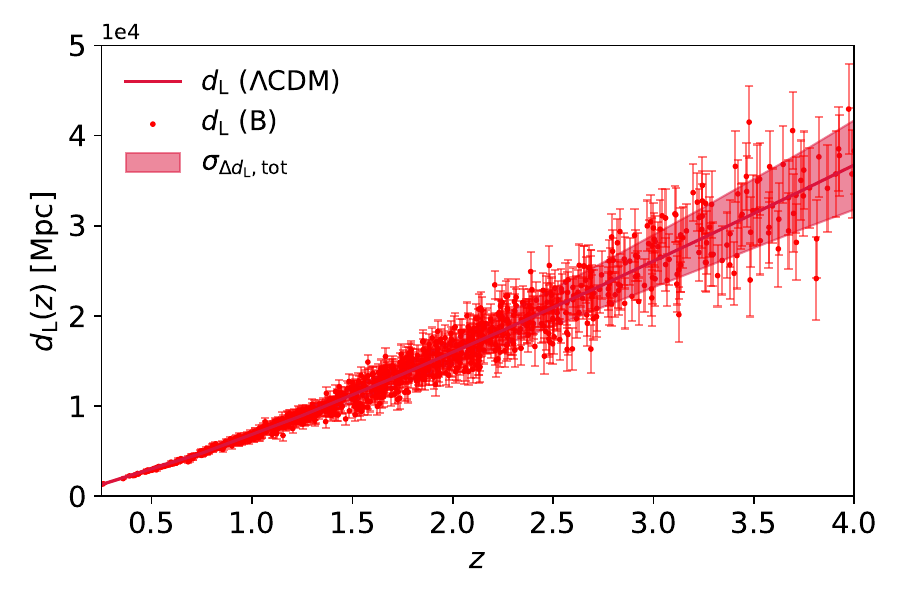}} \\
    \hspace{-0.5cm}
    \subfigure{\includegraphics[width=0.46\textwidth]{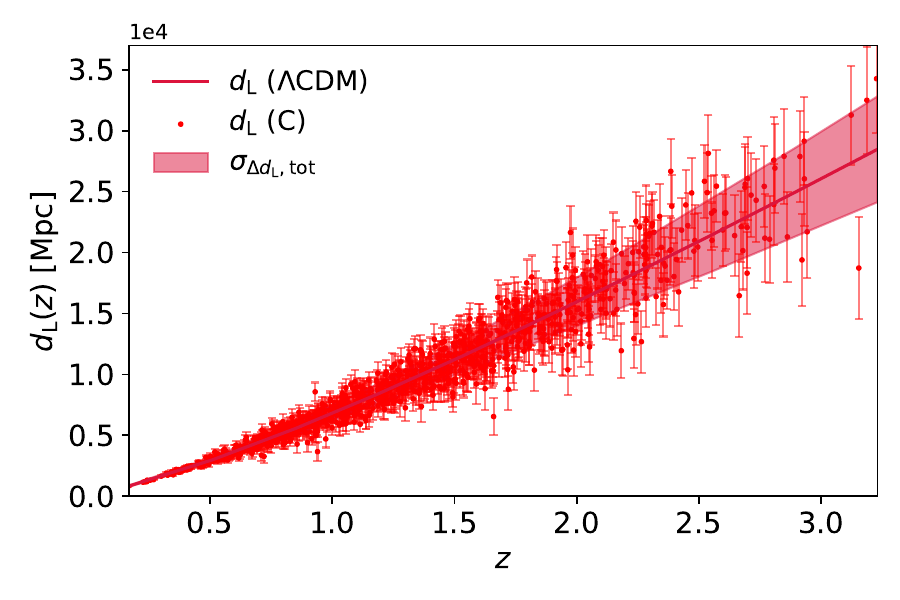}}
    \subfigure{\includegraphics[width=0.46\textwidth]{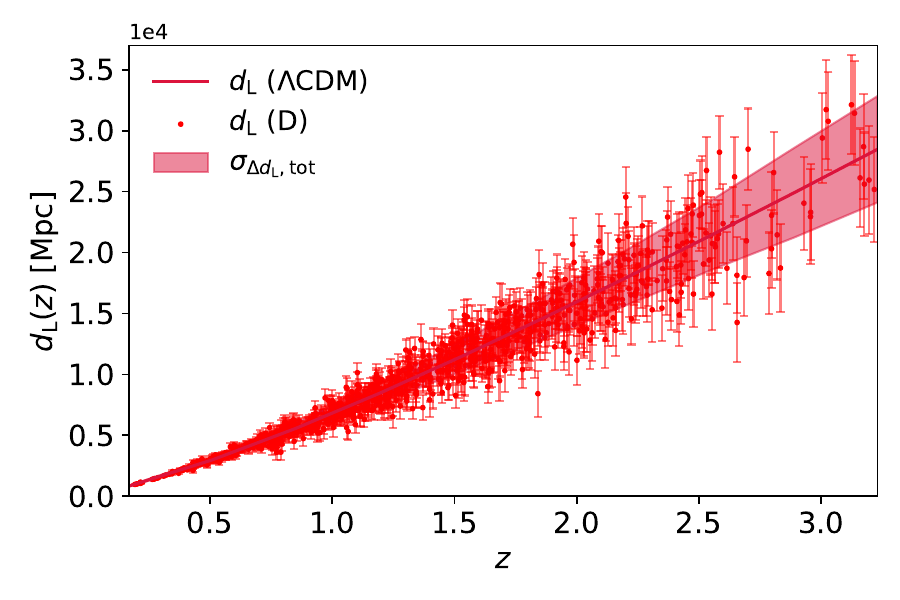}}
    \caption{\reviii{Mock data generated from four simulations. From left to right and top to bottom are the data of catalogs A, B, C, and D, respectively. Data at higher redshifts are rather sparse, hence not shown in this figure. The colored areas represent the average 1 $\sigma$ ranges calculated in redshift bins. Note that the upper limits of $x$-axes are different for single and multiple detectors.}\label{mock data}}
  \end{figure*}

  \begin{figure*}
    \centering
    \hspace{-0.5cm}
    \subfigure{\includegraphics[width=0.45\textwidth]{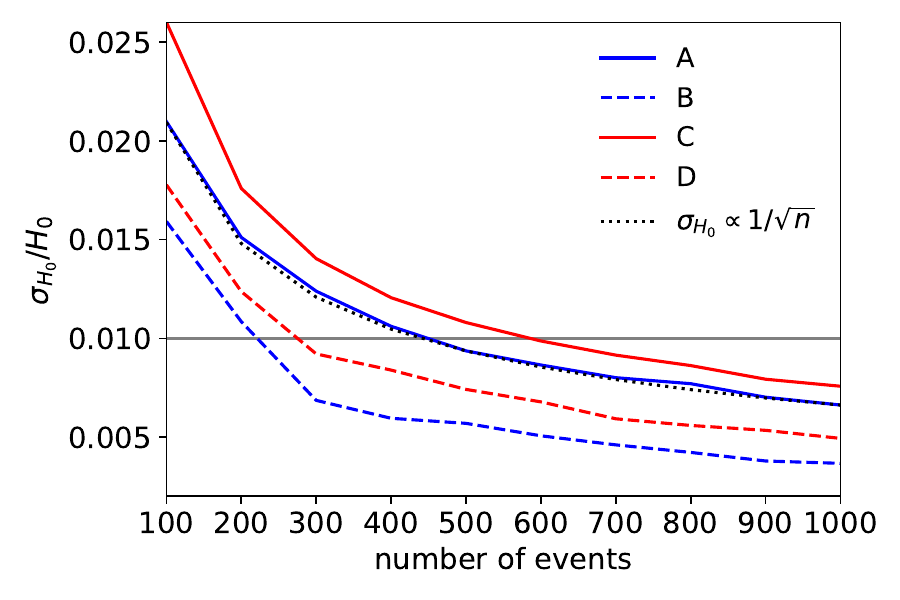}}
    \hspace{-0.2cm}
    \subfigure{\includegraphics[width=0.45\textwidth]{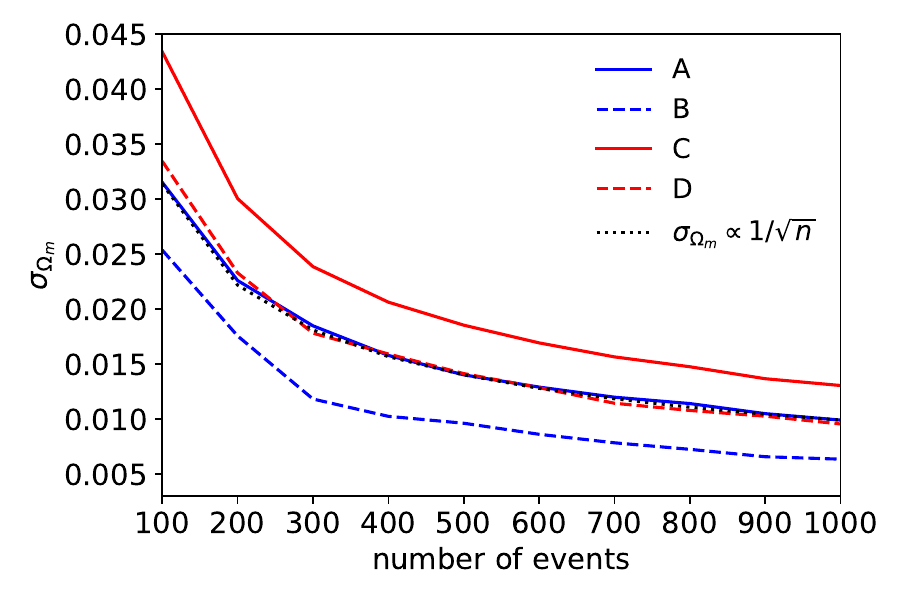}}
    \vspace{-0.3cm}
    \caption{\reviii{Errors of the $\Lambda$CDM parameters $H_0$ (left panel) and
    $\Omega_m$ (right panel). The asymptotic behavior of simulation A is shown with dotted curve in each panel. We set the target fractional accuracy of $H_0$ to $1\%$, which is represented by a grey horizontal line in the left panel.}\label{lcdm error}}
    \vspace{0.2cm}
  \end{figure*}

\section{Results\label{Results}}
\subsection{Simulated catalogs\label{catalogue}}
In the former section, four catalogs of 1,000 GW events were simulated. Shown in \cref{z_distribution} are the redshift distributions of sources, where the curves labeled “theory” are plotted according to \cref{redshift distribution}. It is worth noting
that the histograms of A and B fit the theoretical curve well. Indeed, up to 78.9\%(A) / 82.4\%(B) of the entire samples pass the SNR threshold of ET + CE.

On the other hand, data in catalogs C and D tend to distribute at lower redshifts, since SNR is inversely proportional to $d_{\rm L}$, and the SNR of an event is relatively lower when observed by ET alone. The rates of acceptance drop to 35.7\%(C) / 43.5\%(D).

In addition, the statistical characteristics $(\mu, \sigma)$ of SNR for the four catalogs are: A(22.45, 13.68), B(22.68, 14.16), C(13.08, 8.53), and D(13.48, 8.60). Obviously, using tighter prior on $v$ and more detectors do improve the overall precision.

With these ready-to-use data, we then evaluate the accuracy of distance measurement.

\subsection{Error analysis\label{errorbudget}}

We uniformly divide the redshift range $z \in (0, 5)$ for A and B or $z \in (0, 3.5)$ for C and D into 10 bins, and calculate the average fractional error of $d_{\rm L}$ in each bin. The results are plotted as functions of redshift in \cref{error_z}. Again, it is obvious that smaller $\sigma_v$ and more detectors both lead to smaller errors at given redshifts. Specifically, the improvement of accuracy due to more detectors is evident in the whole redshift range, almost halving the uncertainty in each bin. While the impact of smaller $\sigma_v$ turns out to be more evident at low redshifts. This tendency can be explained by the simplified expression \cref{simplified fisher}. At low redshifts, the average SNR is relatively high; thus, the error induced by $\sigma_v$ overwhelms the $1/\rho$ term, making the alteration of $\sigma_v$ more significant.

\reviii{Besides, by means of simulation,~\cite{Grimm:2020ivq} found that without the help of the EM counterpart, the average relative error of $d_{\rm L}$ in $z \in (0, 1)$ was 0.48 for ET and 0.33 for the ET + LISA + B-DECIGO network, whereas, in the presence of SGRB, we have $\sigma_{\Delta d_{\rm L}} / d_{\rm L} < 0.1$ in the same range for each configuration. This is direct evidence of the benefit of the EM counterpart.}

With dashed curve, we also plot $\sigma_{\Delta d_{\rm L},0} / d_{\rm L}$ in the same figure for comparison. In each case, adopting $\sigma_{\Delta d_{\rm L},0}$ causes overestimation, meaning that it is a rather conservative choice for the error analysis.

\reviii{Another factor usually accounted for is the systematic error due to weak lensing~\cite{Tamanini:2016zlh, 2010PhRvD..81l4046H}, which is a major source of error on $d_{\rm L}$ for high-redshift standard
sirens}
\begin{equation}
    \frac{\sigma_{\Delta d_{\rm L}, {\rm lens}}}{d_{\rm L}} = \mathcolorbox{yellow}{0.066}\left[\frac{\mathcolorbox{yellow}{1 - (1 + z)^{-0.25}}}{\mathcolorbox{yellow}{0.25}}\right]^{1.8}.
\end{equation}
Thus, the total error can be expressed as
\begin{equation}
    \sigma_{\Delta d_{\rm L}, {\rm tot}} = \sqrt{\sigma_{\Delta d_{\rm L}, {\rm lens}}^2 \
    + \sigma_{\Delta d_{\rm L}}^2}.
\end{equation}
\reviii{We also plot the lensing errors in~\cref{error_z}. In each panel, the contribution of lensing is comparable to the instrumental noises.}

So far, we have obtained four groups of $\{z, d_{\rm L}, \sigma_{\Delta d_{\rm L}, {\rm tot}}\}$ values. However, these $(z, d_{\rm L})$ pairs strictly satisfy the functional relationship~\cref{dl_lcdm}, which violates the stochastic nature of the universe. Thus, we will generate the luminosity distances by Gaussian distribution $N(d_{\rm L}, \sigma_{\Delta d_{\rm L}, {\rm tot}})$, which is a good approximation when the error is not too large. The data points as well as associated errors for the four configurations are shown in \cref{mock data}.

\subsection{$\Lambda$CDM cosmology}
\begin{figure*}
  \centering
  \subfigure{\includegraphics[width=0.42\textwidth]{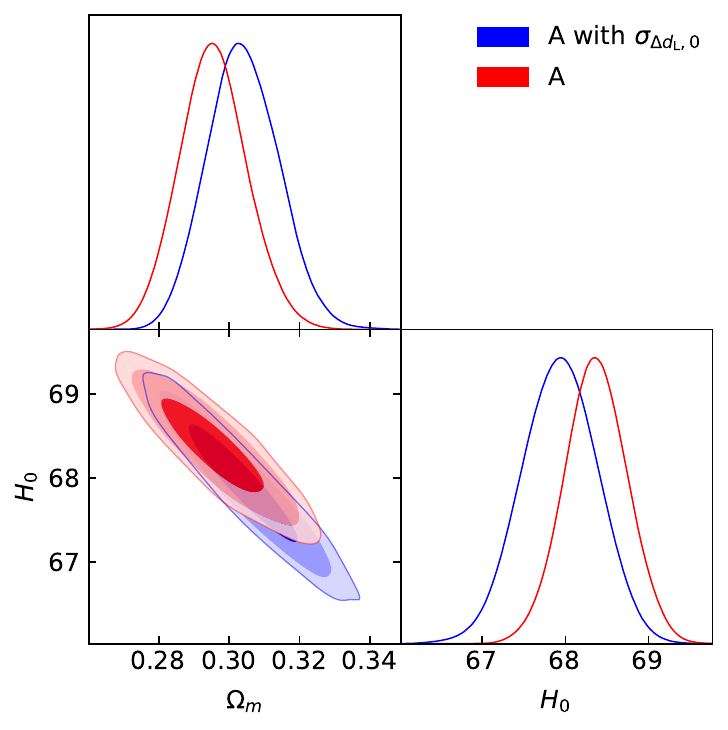}}
  \hspace{0.3cm}
  \subfigure{\includegraphics[width=0.42\textwidth]{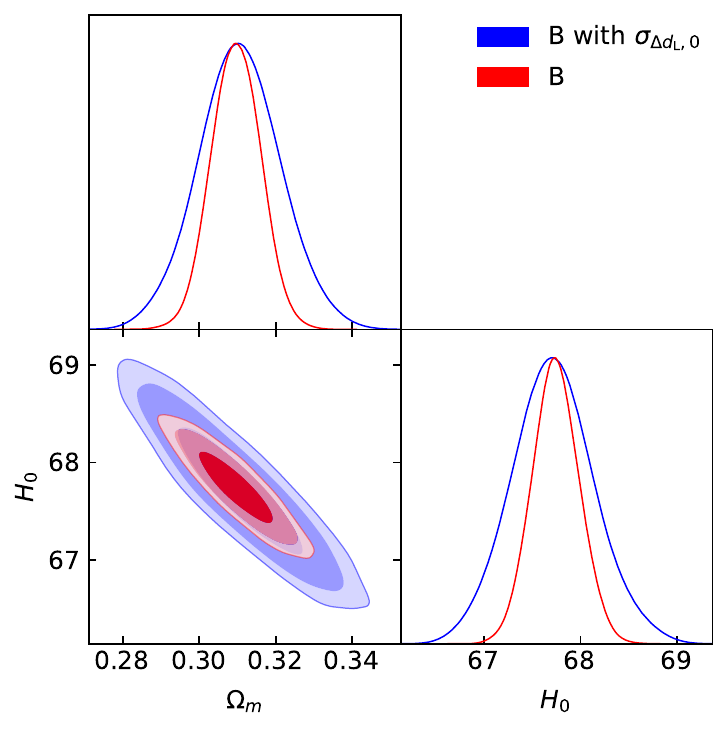}} \\
  \vspace{-0.2cm}
  \subfigure{\includegraphics[width=0.42\textwidth]{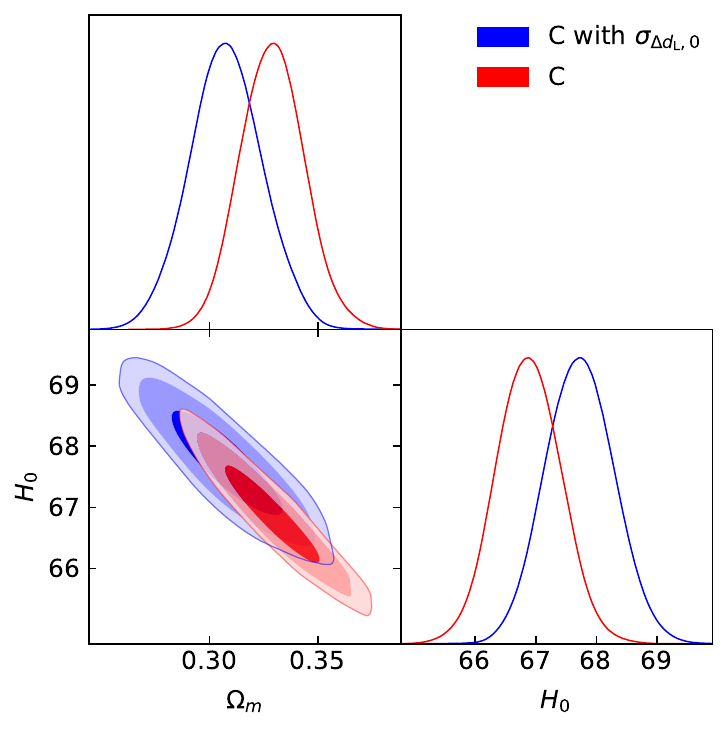}}
  \hspace{0.3cm}
  \subfigure{\includegraphics[width=0.42\textwidth]{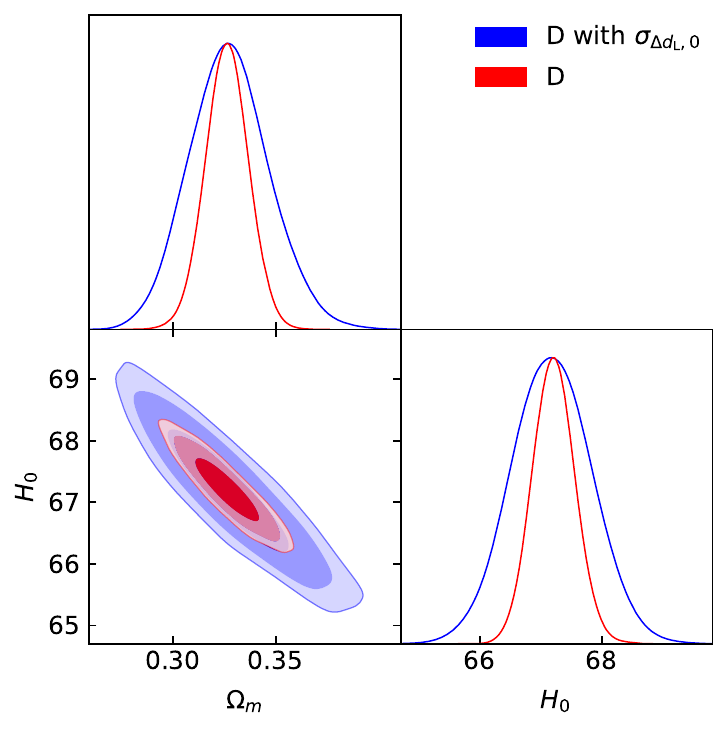}}
  \caption{\reviii{Posterior distributions of the cosmological parameters. The subfigures are arranged in the same order as \cref{mock data}. Shown with contours are the 1 $\sigma$ and 2 $\sigma$ ranges of parameters, whereas marginal distributions are plotted with curves. The red (blue) contours correspond to simulation results (adopting $\sigma_{\Delta d_{\rm L},0}$).}  \label{mcmc}}
\end{figure*}

\begin{table*}
  \renewcommand\arraystretch{1.5}
  \begin{center}
  \begin{tabular}{p{2.5cm}p{2.3cm}<{\centering}p{2.5cm}<{\centering}p{2.3cm}<{\centering}p{2.3cm}<{\centering}}
  \hline
               & A & B & C & D\\
  \hline
  $H_0$                              & $68.37\pm 0.38              $ & $67.74\pm 0.24             $ & $66.88\pm 0.53             $ & $67.21\pm 0.35             $ \\
  $\Omega_m$                         & $0.2955\pm 0.0097             $ & $0.3096\pm 0.0065$ & $0.329\pm 0.015   $ & $0.326\pm 0.011   $ \\
  $H_0$ (with $\sigma_{\Delta d_{\rm L},0}$)       & $67.93\pm 0.45            $ & $67.71\pm 0.43             $ & $67.72\pm 0.56             $ & $67.17\pm 0.65             $ \\
  $\Omega_m$ (with $\sigma_{\Delta d_{\rm L},0}$)  & $0.304\pm 0.010           $ & $0.311\pm 0.011          $ & $0.308 \pm 0.017            $ & $0.327\pm 0.020     $ \\
  \hline
  \end{tabular}
  \end{center}
  \caption{\reviii{1$\sigma$ $(68\%)$ ranges of $\Lambda$CDM parameters constrained with the Markov chain Monte Carlo (MCMC) method. Results in the second and third rows are obtained from simulation, whereas the last two rows show the results obtained by adopting   $\sigma_{\Delta d_{\rm L},0}$.} \label{mcmc table}}
\end{table*}

To determine how many events are required to achieve acceptable precision for cosmological parameters, again we make use of the FIM. Under flat $\Lambda$CDM model, the FIM of parameters $\bm{p} = (H_0, \Omega_m)$ is as follows:
\begin{equation}
    \Gamma^{\Lambda {\rm CDM}}_{ij} = \sum_n \frac{\partial d_{\rm L}(z_n)}{\partial p_i} \
    \frac{\partial d_{\rm L}(z_n)}{\partial p_j} \sigma^{-2}_{\Delta d_{\rm L}, {\rm tot}, n},
\end{equation}
where the partial derivatives are calculated according to \cref{dl_lcdm}, and the covariant matrix is $\Sigma^{\Lambda {\rm CDM}} = \left(\Gamma^{\Lambda {\rm CDM}}\right)^{-1}$. Therefore, $\sigma_{H_0} = \sqrt{\Sigma^{\Lambda {\rm CDM}}_{11}}$ and $\sigma_{\Omega_m} = \sqrt{\Sigma^{\Lambda {\rm CDM}}_{22}}$. For each configuration, we randomly select $100N$ ($N = 1, 2, ..., 10$) events to compose 10 subsets, and these subsets are used in sequence to compute 10 FIMs.

The errors of parameters are presented in \cref{lcdm error}. Both $\sigma_{H_0}$ and $\sigma_{\Omega_m}$ scale roughly as $1/\sqrt{N}$. For each parameter, the improvement in accuracy from large $\sigma_v$ to small $\sigma_v$ is significant , while for $H_0$, the influence of detector number seems moderate at given $N$. As a matter of fact, the detecting capability of ET is poorer than that of the ET + CE network. Judging from the rates of acceptance, the number of events recognized by ET is only about half of ET + CE in the same detecting period. Thus, the values of $N$ should be different when it comes to the comparison between single and multiple detectors (e.g., $N = 2$ for ET and $N = 4$ for ET + CE, with $\sigma_v$ fixed), and it turns out that the impact of detector number is even more significant than that of  $\sigma_v$ value. Analyses in the preceding sections are not affected by this factor since only the individual and average errors are involved there.

To achieve $1\%$ accuracy of $H_0$, the numbers of events required for the four configurations are about \reviii{500, 200, 600, and 300}, respectively. With all 1,000 events, $H_0$ can be constrained to \reviii{$0.66\%$, $0.37\%$, $0.76\%$, and $0.49\%$}, while the errors of $\Omega_m$ are \reviii{0.010, 0.006, 0.013, and 0.010}, respectively. \reviii{These results are comparable to those of~\cite{Maggiore:2019uih}, which declared that a subpercent level accuracy on $H_0$ could be reached by ET.}
\begin{figure*}
  \centering
  \subfigure{\includegraphics[width=0.45\textwidth]{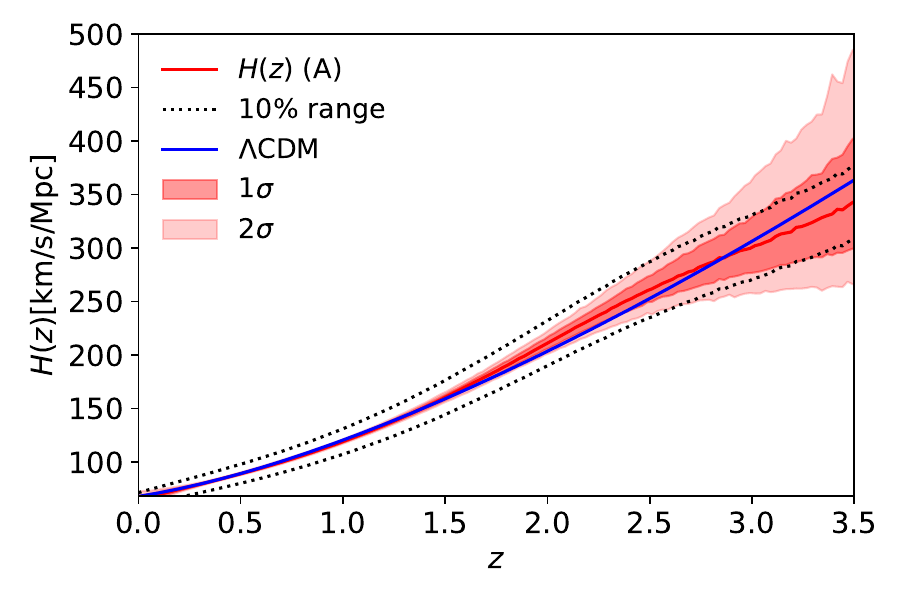}}
  \subfigure{\includegraphics[width=0.45\textwidth]{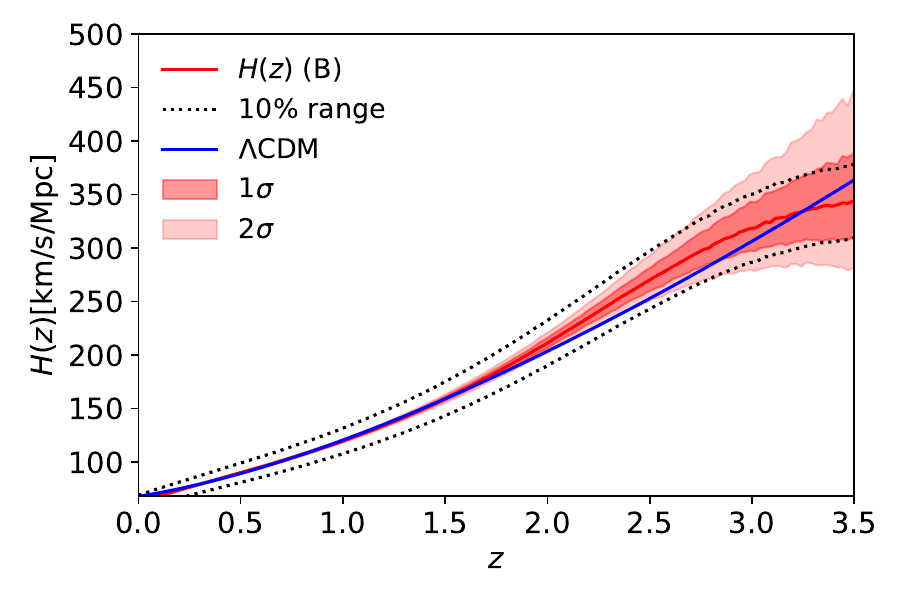}} \\
  \vspace{-0.5cm}
  \subfigure{\includegraphics[width=0.45\textwidth]{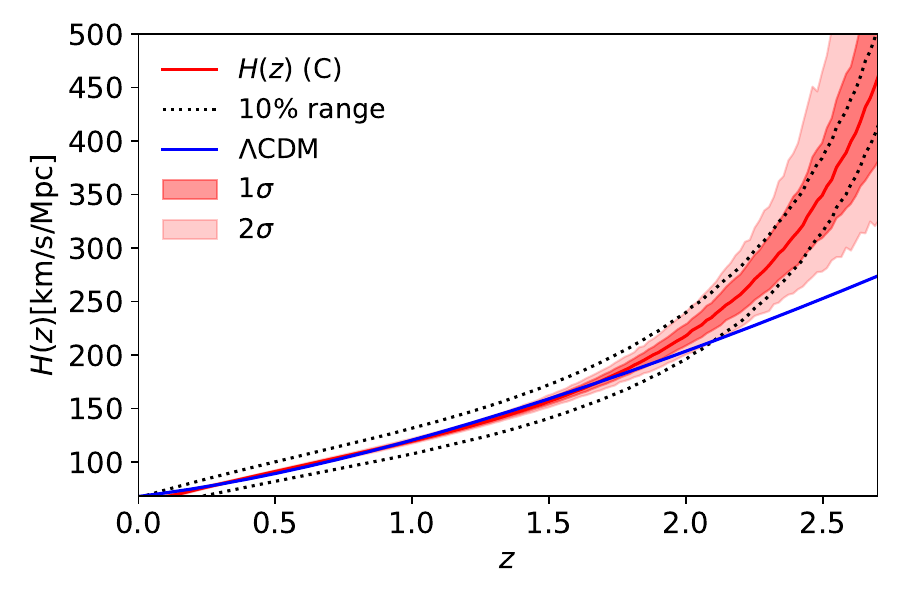}}
  \subfigure{\includegraphics[width=0.45\textwidth]{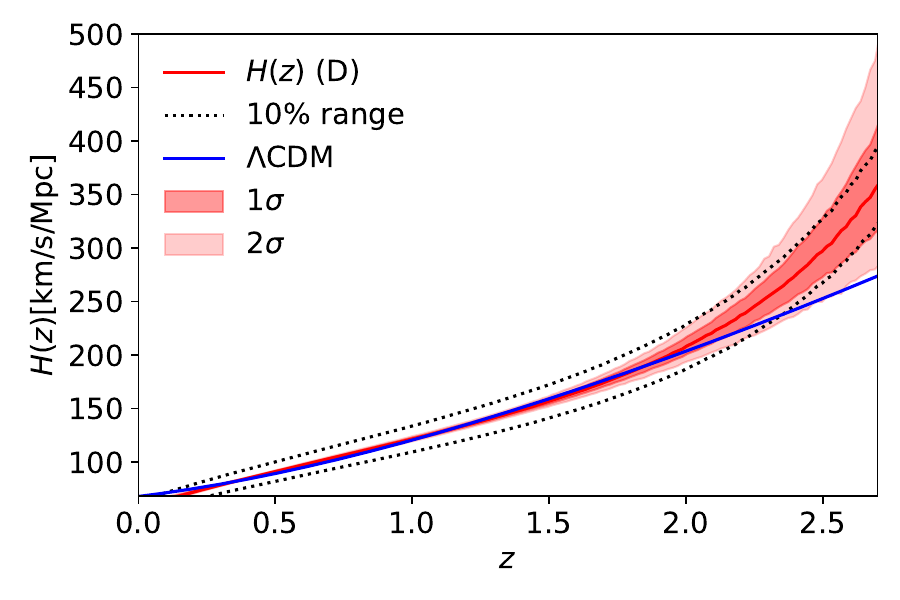}}
  \vspace{-0.5cm}
  \caption{\reviii{Results of Gaussian process for simulations A, B, C, and D, arranged
  in the same order as \cref{mock data}. $10\%$ ranges around $H(z)$
  are plotted to check the detection depth of third-generation gravitational wave detectors. The blue lines correspond to the fiducial $\Lambda$CDM model.} \label{GP}}
\end{figure*}

As supplement, we also present the results of full Bayesian analysis, performed with the public code CosmoMC~\cite{Lewis:2002ah}. \cref{mcmc} and \cref{mcmc table} show the posterior distributions of parameters $\bm{p} = (H_0, \Omega_m)$ constrained from four catalogs of 1,000 events. The blue contours correspond to the results obtained by adopting $\sigma_{\Delta d_{\rm L},0}$. After simple calculation, it is easy to verify that the results of Markov chain Monte Carlo (MCMC) is compatible to those of FIM. And very reasonably, using $\sigma_{\Delta d_{\rm L},0}$ leads to overestimation, regardless of the number of detectors or the value of $\sigma_v$, consistent with our analysis in \cref{errorbudget}.

\subsection{Cosmology with generic $H(z)$}
Despite model-independent approaches, GP~\cite{GAPP}, as a non-parametrization method, can extract information from observational data without any hypothetic functional form. Therefore, GP provides a means to study how accurately a generic function $H(z)$ can be evaluated from data. To this end, we first calculate $d_{\rm L}(z)$ as a GP
\begin{equation}
  d_{\rm L}(z) \sim \mathcal{GP}\left[\mu(z), k(z, z^\prime)\right],
\end{equation}
where $\mu(z)$ and $k(z, z^\prime)$ are the mean value and covariance function, respectively. Then, $H(z)$ can be obtained via
\begin{equation}\label{eq:HfromD}
    H(z) = \frac{c(1 + z)^2}{d_{\rm L}^\prime(z)(1 + z) - d_{\rm L}(z)},
\end{equation}
where “$\prime$” denotes the derivative of redshift. The only assumption behind~\cref{eq:HfromD} is that the geometry of the universe is depicted by the flat Friedmann--Robertson--Walker metric.

Using the open-source code Gaussian Processes in Python (GAPP), we reconstruct $H(z)$ from the four catalogs, and the resulting mean values and errors up to $2 \sigma$ are presented in \cref{GP}. Future GWSS has the advantage of deeper detection depth than conventional standard candles, such as SN Ia. For example, the recently released Pantheon sample~\cite{Riess:2017lxs} consists of 1,048 SNe Ia with maximum redshift 2.26, and most of them are within the range of $z < 1$. While, as is shown in~\cref{GP}, GWSS can constrain $H(z)$ to considerable accuracy at relatively high redshifts. The $1 \sigma$ precision of $10\%$ can be maintained to redshift \reviii{3.12, 3.34, 2.37, and 2.53} by configurations A, B, C, and D, respectively. \reviii{Besides, the $H(z)$ functions reconstructed from A and B agree with the fiducial model (blue curves), even when $z \rightarrow 3.5$, whereas, for C and D, $\Lambda$CDM exceeds the $1\sigma$ ranges at $z = 1.90$ and $z = 2.15$.} In conclusion, adopting multiple detectors and tighter prior both contribute to a precise and unbiased measurement.

{\revi
\begin{figurehere}
  \centering
  \hspace{-0.6cm}
  \includegraphics[width=0.43\textwidth]{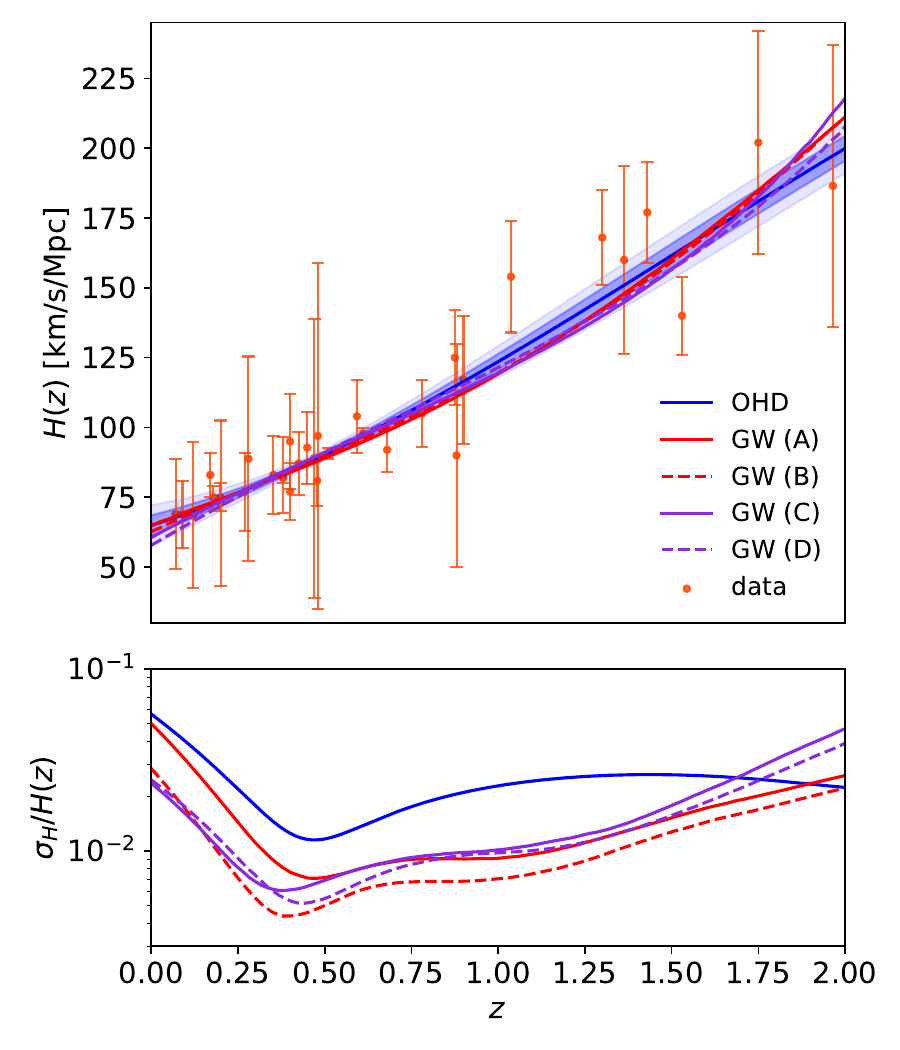}
  \vspace{-0.3cm}
  \caption{The observational Hubble data and reconstructed $H(z)$ function, compared with the results of gravitational waves. Upper panel: observational Hubble data and mean values of $H(z)$. Lower panel: relative errors of $H(z)$. \label{fig:GWvsOHD}}
\end{figurehere}

The constraining ability of future GW data can be better demonstrated when compared with current observations. Compiled by~\cite{Yu:2017iju} are the Hubble parameters measured at different redshifts within $z \in (0, 2)$ (observational Hubble data, OHD) via methods such as the cosmic chronometer and baryon acoustic oscillation. The OHD with error bars and the results of GP are shown in the upper panel of~\cref{fig:GWvsOHD}. For comparison, the mean values and fractional uncertainties of $H(z)$ from GW data and OHD are summarized in the upper and lower panels, respectively. In almost the whole redshift range, $H(z)$ can be measured more accurately from each GW catalog than from OHD.}

{\revi \section{Further comparisons}\label{sec:FurtherComparisons}
The entire pipeline of simulation and data processing is based on several models and hypotheses, such as the distribution of inclination angle and the SFR model. In this section, we make alterations to them and investigate what difference will be brought about.

\subsection{Varied $\bar{v}$ vs. fixed $\bar{v}$}

In this section, we investigate the impact of varied mean value of $v$ in the prior, i.e., replacing \cref{vprior} with
\begin{equation}\label{eq:new_prior}
    p_v(v) \propto \exp\left[-\frac{\left(v - \bar{v}\right)^2}{2\sigma_v^2}\right],
\end{equation}
$\bar{v}$ being the “true” source parameter. This analysis is performed to simulate a situation where $\bar{v}$ is already determined from the SGRB observation with precision $\sigma_v$; thus, the data of SGRB and GW are employed in a more joint manner. In the simplified regime where there is only one stationary interferometer, the consequence of using~\cref{eq:new_prior} can be derived analytically:
\begin{equation}\label{eq:new_error}
    \frac{\sigma_{\Delta d_{\rm L}}}{d_{\rm L}} = \sqrt{\frac{1}{\rho^2} + K\sigma_v^2}, \quad K = \frac{v_1^2}{1 + v_2^2\sigma_v^2\rho^2}.
\end{equation}
By numerical calculation, it is easy to confirm that $K > 1$ for the typical values of $v_1$, $v_2$, and $\rho$; thus, varying $\bar{v}$ usually increases error compared with~\cref{simplified fisher}. It should be noted that~\cref{simplified fisher} is estimated in the face-on limit, while~\cref{eq:new_error} is valid whether $v$ is 1 or not; thus, the performances of these two priors remain to be compared through simulation. Besides, in practice, the uncertainties of $v$ constrained from SGRB data vary among events; thus, using a single value $\sigma_v$ is a rather rough choice.

Following the prescription of error budget given in~\cref{errorbudget}, we apply~\cref{eq:new_prior} to the simulated GW catalogs, and the instrumental errors are shown in~\cref{fig:error_varied_v} as solid curves. For comparison, we also plot the results of~\cref{vprior} in the same figure with dotted curves. It is obvious that varying $\bar{v}$ increases the errors of $d_{\rm L}$. Besides, for the cosmological parameters $\bm{p} = \{H_0, \Omega_m\}$, we have $\sigma_{H_0} / H_0 =$ 0.74\%, 0.37\%, 0.84\%, 0.51\% and $\sigma_{\Omega_m} =$ 0.011, 0.006, 0.014, 0.010, respectively, which are also slightly larger than those of fixed $\bar{v}$. Moreover, the influence of adopting~\cref{eq:new_error} is more evident when $\sigma_v = 0.05$ since $v$ is allowed to vary in a wider range in this case.

\begin{figurehere}
  \centering
  \vspace{0.3cm}
  \hspace{-0.7cm}
  \includegraphics[width=0.48\textwidth]{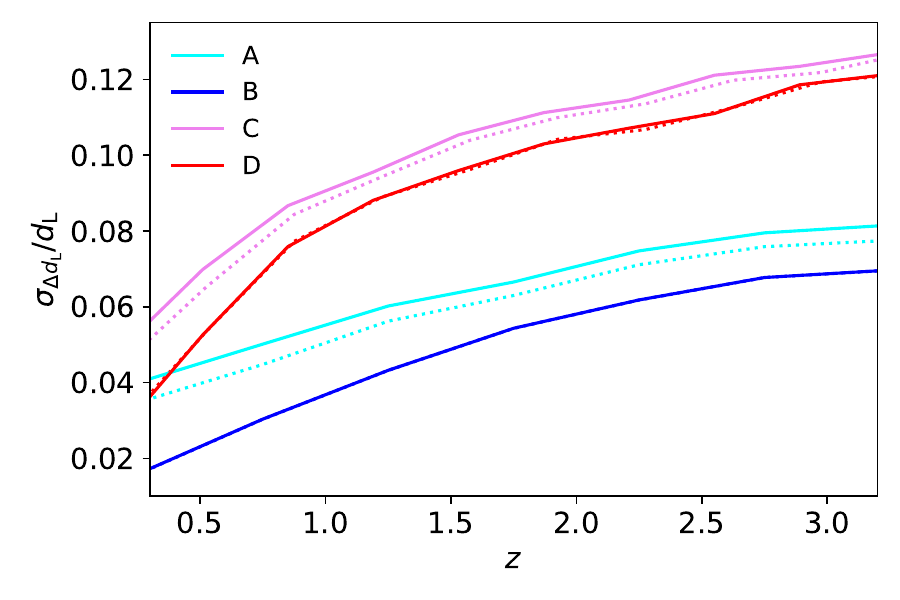}
  \vspace{-0.3cm}
  \caption{The instrumental errors of $d_{\rm L}$ as functions of redshift. The solid and dotted curves are obtained by adopting~\cref{vprior} and~\cref{eq:new_error}, respectively, and the results of the four configurations are plotted with different colors.\label{fig:error_varied_v}}
  \vspace{0.3cm}
\end{figurehere}

\subsection{Other star formation rate models}
The simulation of GW sources is based on the hypothesis that the NS formation follows the SFR. While, as a matter of fact, the high-redshift SFR is uncertain at present. Pioneering studies~\cite{Robertson_2011, Wang:2013yrw, Kistler:2013jza} have revealed that GRBs observed by Swift provide a biased measurement of the SFR history, with an enhancement of $\sim (1 + z)^{0.5}$. \cite{Robertson_2011} reported that this factor could be readily explained if GRBs occur primarily in low-metallicity galaxies proportionally more numerous at earlier times. However, after considering this trend, the SFR beyond $z = 4$ derived from GRBs is still much higher than that from other surveys, such as ultraviolet (UV) and far-infrared (FIR) measurements. \cite{Wang:2013yrw} provided an explanation for the high-redshift GRB rate excess by considering the GRBs produced by rapidly rotating metal-poor stars from low masses.

It is beyond the scope of this paper to explore the physical origins of the discrepancies between different SFR surveys. On the contrary, apart from Fiducial + PopIII (\cref{eq:SFR}), we adopt two additional SFR models, one inferred from GRBs~\cite{Kistler:2013jza} (dubbed Kistler):
\begin{equation}
    {\rm SFR}(z) \propto \left[(1 + z)^{a\zeta} + \left(\frac{1 + z}{B}\right)^{b\zeta} + \left(\frac{1 + z}{C}\right)^{c\zeta}\right]^{1/\zeta},
\end{equation}
with $\{a, b, c, B, C, \zeta\} = \{3.4, -0.3, -2.5, 5160, 11.5, -10\}$, and the other from UV and FIR measurements~\cite{Hopkins:2006bw, Wang:2013yrw} (dubbed Hopkins):
\begin{equation}
    {\rm SFR}(z) \propto 
        \begin{cases}
        (1 + z)^{3.44}, & z < 0.97,\\
        (1 + z)^{-0.26}, & 0.97 < z < 4.48,\\
        (1 + z)^{-7.8}, & 4.48 < z.
        \end{cases}
\end{equation}

The simulated GW catalogs are compared in~\cref{fig:SFR_comparison}. For clarity, we only present the events obtained under configuration B. The red, blue, and black histograms represent the distributions of GW sources generated from Hopkins, Kistler, and Fiducial + PopIII, respectively, and curves with the same colors correspond to the theoretical BNS merger rates. For both Hopkins and Kistler, the peaks of $P(z)$ come at lower redshifts ($z = 1.1$) than Fiducial + PopIII ($z = 1.5$), and the high-redshift merger rates are slightly larger, especially for Kistler, due to the so-called “GRB rate excess” phenomenon.

Under configuration B, the uncertainties of cosmological parameters are $\sigma_{H_0} / H_0 =$ 0.295\%, 0.349\%, 0.367\% and $\sigma_{\Omega_m} = $ 0.0057, 0.0064, 0.0064, for Hopkins, Kistler, and Fiducial + PopIII, respectively. The errors of each parameter are in the same order of magnitude. The comparison between simulations agrees with intuition, i.e., parameters can be better constrained from catalogs with more low-redshift (high-SNR) events.

\begin{figurehere}
  \centering
  \includegraphics[width=0.48\textwidth]{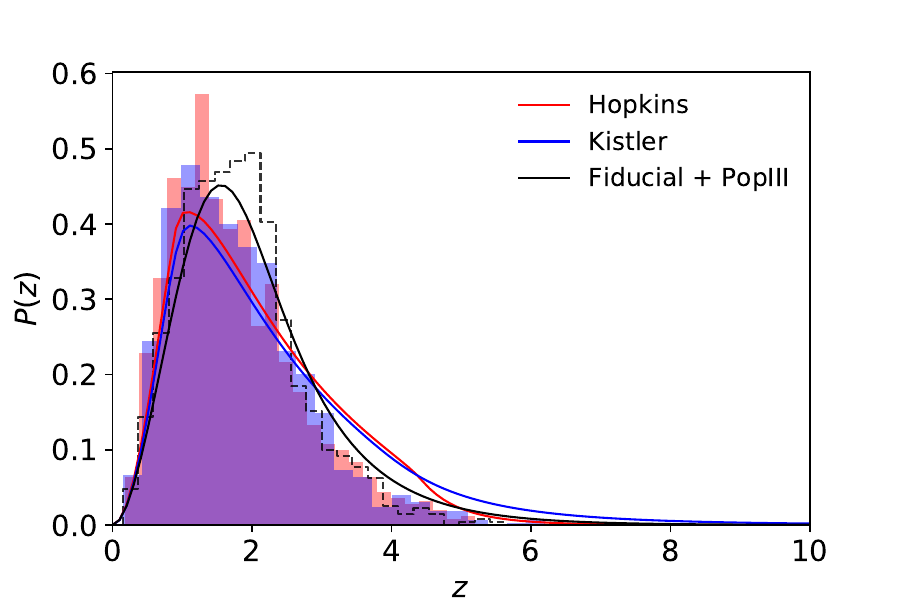}
  \caption{The redshift distributions of gravitational wave events (configuration B) simulated according to different star formation rate models. The red, blue, and black histograms represent the results of SFR models, Hopkins, Kistler, and Fiducial + PopIII, respectively, and the curves with the same colors correspond to the theoretical BNS merger rates.\label{fig:SFR_comparison}}
  \vspace{0.4cm}
\end{figurehere}}

{\revii \subsection{Other short gamma-ray burst models}
In former sections, we have assumed that SGRB follows a Gaussian distribution peaked at $v = 1$, in concordance with~\cite{Nissanke:2009kt, PhysRevD.74.063006}. The modeling of SGRB plays a crucial role in the process of parameter estimation, and several models, such as the jet and cocoon scenarios~\cite{10.1093/mnrasl/sly061}, have been proposed to describe the SGRB features in the afterglow of GW. Therefore, we discuss this issue by comparing our model with the Gaussian structured jet profile
\vspace{-0.2cm}
\begin{equation}\label{iota_prior}
    p_\iota(\iota) \propto \left[-\frac{\iota^2}{2\sigma_\iota^2}\right],
\end{equation}
which is a Gaussian distribution on $\iota$ with standard deviation $\sigma_\iota$. \cite{10.1093/mnrasl/sly061} constrained $\sigma_\iota = 0.091^{+0.037}_{-0.040}$ rad, $0.057^{+0.025}_{-0.023}$ rad, and $0.076^{+0.026}_{-0.027}$ rad from different data combinations, whereas~\cite{Howell:2018nhu} reported $\sigma_\iota = 4.7^{+1.1}_{-1.1}$ deg. These values are very close to our choice of $\sigma_v = 0.003$; thus, we adopt $\sigma_\iota = \cos^{-1}(1 - 0.003) = 0.077$, replace~\cref{vprior} by~\cref{iota_prior}, and repeat simulations B and D.

\begin{figurehere}
  \centering
  \vspace{0.5cm}
  \hspace{-0.5cm}
  \includegraphics[width=0.45\textwidth]{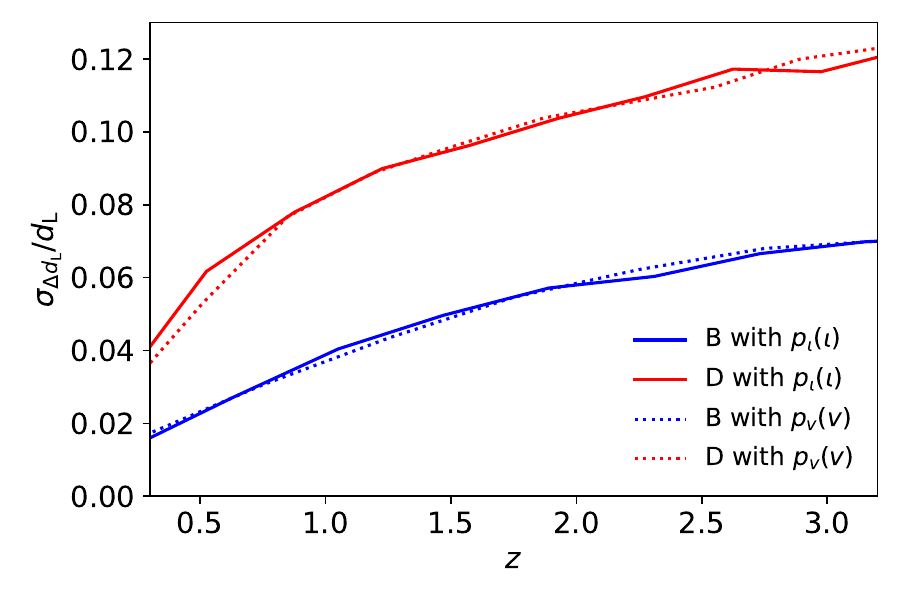}
  \vspace{-0.3cm}
  \caption{Relative errors of $d_{\rm L}$ as functions of redshift. Results obtained by adopting $p_\iota(\iota)$ (or $p_v(v)$) are plotted with solid (or dotted) curves, and blue (or red) curves correspond to configuration B (or D).\label{fig:error_v_iota_BD}}
  \vspace{0.4cm}
\end{figurehere}

It appears that the instrumental errors $\sigma_{\Delta d_{\rm L}} / d_{\rm L}(z)$ of different priors are indistinguishable~(\cref{fig:error_v_iota_BD}). Thus, we further calculate the uncertainties of $\Lambda$CDM parameters. With~\cref{iota_prior}, we obtain $\sigma_{H_0} / H_0 =$ 0.34\%, 0.58\% and $\sigma_{\Omega_m} =$ 0.006, 0.011 for B, D, respectively. The results of adopting $p_\iota(\iota)$ is only barely different from those of $p_v(v)$ since they both indicate that $\iota$ is confined in a small range ($< 5$ deg) around 0.}

\section{Conclusion\label{Conclusion}}
In this paper, we address the problem of degeneracy between luminosity distance and inclination angle in the estimation of GW parameters, especially near the face-on limit.
\reviii{For GW-SGRB events, the sky locations of sources can be pinpointed with techniques such as identifying the host galaxies, and the mass parameters can be determined with exquisite accuracy from the phase of GW; thus, $d_{\rm L}$ and $v = \cos \iota$ are treated as independent of other parameters.}
We first calculate the FIM of these two parameters and then consider a Gaussian prior on the inclination based on assumed knowledge about SGRB. This pipeline is tested via simulation. Four catalogs, each with 1,000 events detected by 3G GW detectors, are simulated, and the settings are \{A: ET + CE /$\ \sigma_v = 0.05$, B: ET + CE /$\ \sigma_v = 0.003$, C: ET /$\ \sigma_v = 0.05$, D: ET /$\ \sigma_v = 0.003$\}. Although future SGRB observations may give $\sigma_v$ other than these two values, our research may give a clue to the changes in results when $\sigma_v$ varies.

It turns out that the detector network, ET + CE, can recognize more and farther events than a single detector ET in the same detecting period.
By analyzing the error of $d_{\rm L}$, we find that using more detectors and tighter prior on the inclination can both improve the precison of measurement. Also considered is the performance of a widely adopted formula $\sigma_{\Delta d_{\rm L},0} = 2d_{\rm L} / \rho$, which overestimates the error of $d_{\rm L}$ for each configuration.

The simulated catalogs are further applied to constrain cosmological parameters. For $\Lambda$CDM cosmology, \reviii{500, 200, 600, and 300} events are required for the four configurations to achieve $1\%$ $H_0$ accuracy. With all 1,000 events, $H_0$ is constrained to \reviii{$0.66\%$, $0.37\%$, $0.76\%$, and $0.49\%$}, whereas the errors of $\Omega_m$ are \reviii{0.010, 0.006, 0.013, and 0.010}, respectively. The MCMC method produces results compatible with those of FIM. Regarding the Hubble parameter as a free function of redshift and using the GP method, we find that $H(z)$ can be measured with uncertainties less than $10\%$ up to redshifts \reviii{3.12, 3.34, 2.37, and 2.53}, indicating that future GWSS can probe cosmology at higher redshift than the observed standard candles to date. The advantage of future GWSS over current OHD is also obvious.
Besides, adopting more detectors and tighter prior both contribute to a precise and unbiased result.

{\revi In addition, we have also considered the effect of altering some of the formulae and models used in the simulation and data processing, such as SFR and the distribution of inclination angle. Detailed analyses can be found in~\cref{sec:FurtherComparisons}.}

In this paper, the prior distributions of inclination are based only on limited observations. With upcoming detections on SGRB and GW, as well as more reliable estimations of the joint event rate, we expect that more rigorous results can be achieved.

\Acknowledgements{The research of this article is supported by the National Natural Science Foundation of China (Grant No. 11675032, 12075042). We thank Weiqiang Yang of Liaoning Normal University for providing the modified CosmoMC code.}

\InterestConflict{The authors declare that they have no conflict of interest.}


\providecommand{\href}[2]{#2}\begingroup\raggedright\endgroup

\end{multicols}
\end{document}